\documentclass[usenatbib]{mnras}
\usepackage{textcomp}
\usepackage[utf8]{inputenc}
\usepackage{lscape}
\usepackage{lastpage}
\usepackage{amsmath,amssymb}
\usepackage{mathrsfs}	
\usepackage{multicol}
\usepackage[dvipdfmx]{graphicx}
\usepackage{caption}
\usepackage{subcaption}
\usepackage{multirow}
\graphicspath{{figures/}}

\usepackage{times}
\usepackage{rotating}
\catcode`\@=11
\def\gta{\ifmmode{\,\mathrel{\mathpalette\@versim>\,}}
    \else{$\,\mathrel{\mathpalette\@versim>}\,$}\fi}
\def\lta{\ifmmode{\,\mathrel{\mathpalette\@versim<\,}}
    \else{$\,\mathrel{\mathpalette\@versim<}\,$}\fi}
\def\@versim#1#2{\lower 2.9truept \vbox{\baselineskip 0pt \lineskip
    0.5truept \ialign{$\m@th#1\hfil##\hfil$\crcr#2\crcr\sim\crcr}}}
\catcode`\@=12

{\newif\ifnotend
\notendtrue
\def\veclist{ABCDEFGHIJKLMNOPQRSTUVWXYZabcdefghijklmnopqrstuvwxyz.}
\def\top#1#2.{#1}
\def\tail#1#2.{#2.}
\loop\expandafter\xdef\csname bb\expandafter\top\veclist\endcsname%
{{\noexpand\bf\expandafter\top\veclist}}
\edef\veclist{\expandafter\tail\veclist}
\if\veclist.\notendfalse\fi\ifnotend\repeat}

\newcommand{\Zsun}      {Z_{\odot}}

\newcommand{\Te}        {T_{\rm e}}
\newcommand{\Ao}        {A_0}

\newcommand{\HI}	{{\rm H\textsc{i}}}
\newcommand{\HII}   {{\rm H\textsc{ii}}}
\newcommand{\OH}    {{\rm O/H}}
\newcommand{\logoh} {\log({\rm O/H})}

\newcommand{\magg}      {{\rm m}_g}

\newcommand{\Maggsun}   {{\rm M}_{g,\odot}}

\newcommand{\MLgB}      {\Upsilon_g^{{\rm B}}}

\newcommand{\MLgA}      {\Upsilon_g^{{\rm A}}}

\newcommand{\MLgS}      {\Upsilon_g^{{\rm S1}}}

\newcommand{\xyz}   {(x,y,z)}
\newcommand{\xyzi}  {(x_i,y_i,z_i)}

\newcommand{\vxyz}  {(v_x,v_y,v_z)}
\newcommand{\vxyzi} {(v_{x,i},v_{y,i},v_{z,i})}

\newcommand{\vy}    {v_y}

\newcommand{\xini}  {x_i}
\newcommand{\yini}  {y_i}
\newcommand{\zini}  {z_i}
\newcommand{\rini}  {r_i}
\newcommand{\vini}  {v_i}
\newcommand{\vxini} {v_{x, i}}
\newcommand{\vyini} {v_{y, i}}
\newcommand{\vzini} {v_{z, i}}

\newcommand{\vecri} {{\bf r_i}}
\newcommand{\vecv}  {{\bf v}}
\newcommand{\vc}    {v_c}
\newcommand{\vecvi} {{\bf v_i}}
\newcommand{\Lz}    {L_z}
\newcommand{\thetaddo}  {\theta_{68}}
\newcommand{\thetaR}    {\theta_{\rm B}}

\newcommand{\Ndm}   {N_{\rm dm}}
\newcommand{\Nst}   {N_{\star}}
\newcommand{\Ngas}  {N_{\rm gas}}
\newcommand{\sldm}  {l_{\rm dm}}
\newcommand{\slst}  {l_{\star}}
\newcommand{\slgas} {l_{\rm gas}}
\newcommand{\mpart} {m_{\rm part}}
\newcommand{\mpartgas} {m_{\rm gas}}

\newcommand{\Mi}    {M_1}
\newcommand{\Mii}   {M_2}
\newcommand{\tdf}   {t_{\rm fric}}
\newcommand{\tcross}{t_{\rm cross}}
\newcommand{\rt}    {r_{\rm t}}

\newcommand{\Mdm}       {M_{\rm dm}}
\newcommand{\Mstar}     {M_\star}
\newcommand{\Mgas}      {M_{\rm gas}}
\newcommand{\hstar}     {h_\star}
\newcommand{\zstar}     {z_\star}
\newcommand{\hgas}      {h_{\rm gas}}

\newcommand{\Riiv}      {R_{25}}
\newcommand{\rhodm}     {\rho_{\rm dm}}
\newcommand{\rhostar}   {\rho_\star}
\newcommand{\rhogas}    {\rho_{\rm gas}}
\newcommand{\Sigmagas}  {\Sigma_{\rm gas}}
\newcommand{\sech}      {{\rm sech}}

\newcommand{\Phitot}    {\Phi_{\rm tot}}
\newcommand{\Phidm}     {\Phi_{\rm dm}}

\newcommand{\bbxi}      {{\boldsymbol \xi}}

\newcommand{\tmax}  {t_{\rm max}}
\newcommand{\vrot}  {v_{\rm rot}}

\newcommand{\kpc}   {\,{\rm kpc}}
\newcommand{\Mpc}   {\,{\rm Mpc}}
\newcommand{\Msun}  {\,M_{\odot}}
\newcommand{\Gyr}   {\,{\rm Gyr}}
\newcommand{\Myr}   {\,{\rm Myr}}

\newcommand{\kms}   {\,{\rm km\,s^{-1}}}
\newcommand{\asec}  {\,{\rm arcsec}}
\newcommand{\cm}    {\,{\rm cm}}
\newcommand{\K}     {\,{\rm K}}

\newcommand{\rapex}	{\textquoteright\,}
\newcommand{\lapex}	{\textquoteleft}
\newcommand{\DD}    {\partial}
\newcommand{\dd}    {{\rm d}}

\defcitealias{Cannon2014}   {C14}
\defcitealias{Annibali2016} {A16}
\defcitealias{Annibali2019} {A19}
\defcitealias{Annibali2019b}{A19b}
\defcitealias{Sacchi2016}   {S16}

\begin{document}

\date{}

\pubyear{}

\title[Hydro $N$-body simulations of the XMP galaxy DDO 68]{Dancing in the void: hydrodynamical $N$-body simulations of the extremely metal poor galaxy DDO 68}
{}
\author[R. Pascale et al.]{
R. Pascale$^1$\thanks{E-mail: raffaele.pascale@inaf.it}, 
F. Annibali$^1$,
M. Tosi$^1$, 
F. Marinacci$^2$, 
C. Nipoti$^{1,2}$,
M. Bellazzini$^1$, 
\newauthor
D. Romano$^1$, 
E. Sacchi$^3$,
A. Aloisi$^4$,
M. Cignoni$^{5,6,7}$ 
\\ \\
$^1$INAF - Osservatorio di Astrofisica e Scienza dello Spazio di Bologna, Via Gobetti 93/3, 40129 Bologna, Italy \\ 
$^2$Dipartimento di Fisica e Astronomia \lapex Augusto Righi\rapex, Università di Bologna, via Gobetti 93/2, 40129, Bologna, Italy\\
$^3$ Leibniz-Institut f\"ur Astrophysik Potsdam, An der Sternwarte 16, 14482 Potsdam, Germany \\
$^4$ Space Telescope Science Institute, 3700 San Martin Drive, Baltimore, MD 21218, USA \\
$^5$ Dipartimento di Fisica, Università di Pisa, Largo Bruno Pontecorvo 3, 56127, Pisa, Italy \\
$^6$ INFN, Largo B. Pontecorvo 3, 56127, Pisa, Italy \\
$^7$ Osservatorio astronomico di Capodimonte, Vicolo Moiariello 16, 80131, Napoli, Italy \\}%

\label{firstpage}
\pagerange{\pageref{firstpage}--\pageref{lastpage}}

\maketitle

\begin{abstract}
Using hydrodynamical $N$-body simulations, we show that the observed structure and kinematics of the extremely metal-poor, dwarf irregular galaxy DDO 68 is compatible with a merger event with at least two smaller satellite galaxies. We were able to obtain a self-consistent model that simultaneously reproduces several of its observed features, including: the very asymmetric and disturbed shape of the stellar component, the overall $\HI$ distribution and its velocity field, the arc-like stellar structure to the west, the low-surface brightness stellar stream to the north. The model implies the interaction of the main progenitor of DDO 68 with two systems with dynamical masses $7\times10^8\Msun$ and almost $10^8\Msun$ -- 1/20 and 1/150 times the dynamical mass of DDO 68, respectively. We show that the merger between DDO 68 and the most massive of its satellites offers a route to explain the large offset of DDO 68 from the mass-metallicity relation. Assuming that the interacting galaxies have metallicities prior to the merger compatible with those of galaxies with similar stellar masses, we provide quantitative evidence that gas mixing alone does not suffice at diluting the gas of the two components; according to our simulations, the $\HII$ regions observed along the Cometary Tail trace the low metallicity of the accreted satellite rather than that of DDO 68's main body. In this case, the mass corresponding to the low metallicity is that of the secondary body and DDO 68 becomes consistent with the mass-metallicity relation.

\end{abstract}


\maketitle

\section{Introduction}
\label{sec:int}

\begin{figure*}
    \centering
    \includegraphics[width=1.\hsize]{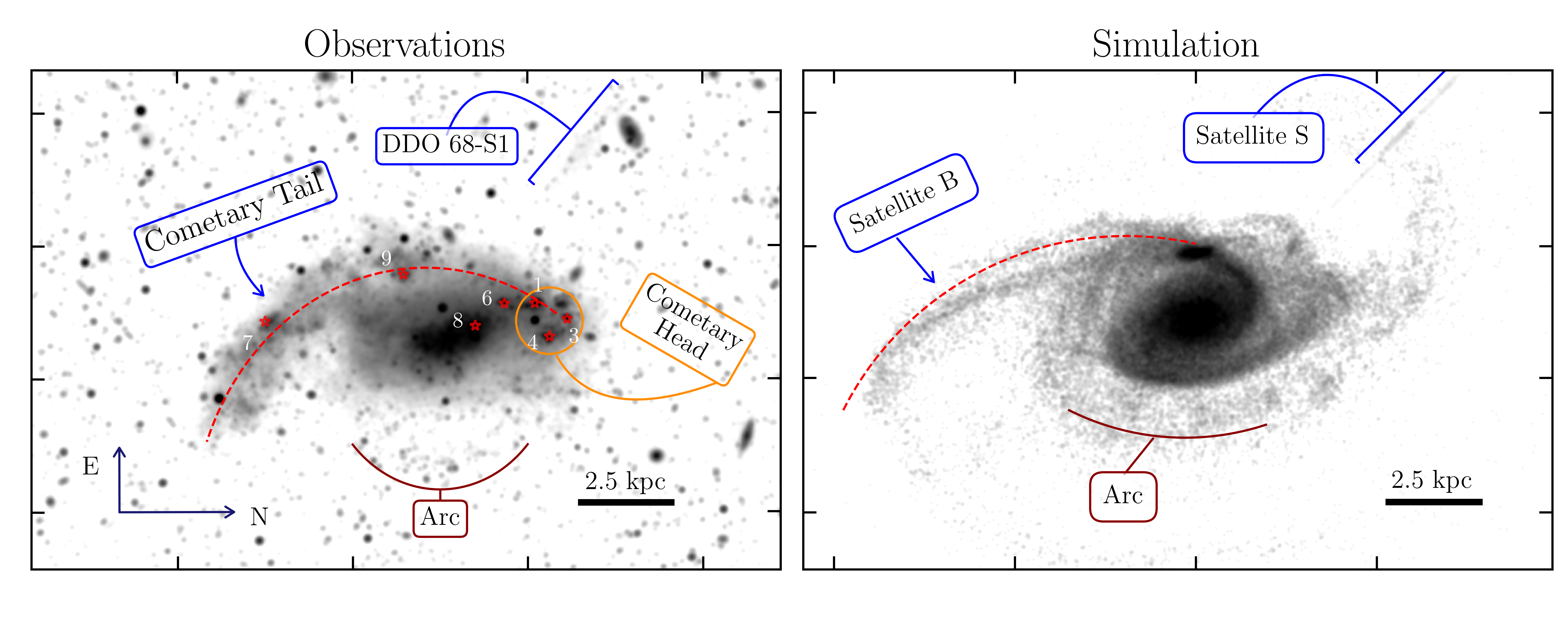}
    \caption{Left panel: LBT g-band image of DDO 68. We show the position of the Cometary Tail (blue arrow), the Cometary Head (yellow circle), where some of the $\HII$ regions are clearly visible as dark spots (see text for details), the position of the stellar stream DDO 68-S1 (blue bracket) and the Arc (dark red arc). All the $\HII$ regions referred in Fig.~\ref{fig:mzr} are shown by red stars. East is up, North is to the right. Right panel: mock observation of DDO 68 from the simulation of Section~\ref{sec:res}. All the features reproduced by the simulation are marked as in the left panel (see Section~\ref{sec:res} for details on similarities and differences between model and data). Both images have the same surface brightness limit (29 mag$\asec^2$), the same spatial resolution (pixel size $0.225\asec\times0.225\asec$) and the same Gaussian smoothing (FWHM $\simeq1.9\asec$). In both panels, we follow DDO 68 B from the Tail to the Head with a dashed red line. The field in each panel is $13.8\times20.7\kpc^2$.}\label{fig:ddo68}
\end{figure*}

With an average oxygen abundance as low as $12+\log(\OH)=7.14\pm0.07$ (e.g. \citealt{Pustilnik2005}; \citealt{Izotov2007}; \citealt[hereafter A19]{Annibali2019}), DDO 68 (UGC 5340), located at a heliocentric distance $D=12.65\Mpc$ \citep[hereafter S16]{Sacchi2016}, ranks itself as one of the most metal-poor dwarf galaxies known to date \citep{Izotov1999b,Pustilnik2005,Pustilnik2008,Izotov2007,Izotov2009b,Izotov2009,Skillman2013,Hirschauer2016,Hsyu2017,Izotov2018,Izotov2019,Kojima2020,Pustilnik2020}. Galaxies with similar record-breaking metallicity, corresponding to less than $1/40\Zsun$ \citep{Caffau2008}, are, for instance, the blue compact dwarf (BCD) galaxies I Zw 18 ($12+\log(\OH)=7.18\pm0.01$; \citealt{Skillman1993,Izotov1999}), Leo P \citep[$12+\log(\OH)=7.17\pm0.04$;][]{Skillman2013}, or the star-forming dwarf galaxy J0811+4730 \citep[$12+\logoh = 6.98\pm0.02$;][]{Izotov2018}.

The existence of the galaxy mass-metallicity relation (MZR) is typically understood as the result of a complex interplay between metal production by star formation, accretion of metal poor gas and metal loss through galactic outflows: more massive galaxies are metal richer because they are more efficient at forming stars and less subject to the dispersion of metals via galactic winds thanks to their deeper gravitational potential well \citep{Lee2006,Brooks2007,Ma2016}. The low metallicity of DDO 68 is peculiar in light of its relatively large stellar mass and luminosity. DDO 68 is, in fact, an outlier of the mass-metallicity relation (e.g. \citealt{Pustilnik2005,Berg2012,McQuinn2020}) since it is eXtremely Metal Poor (XMP, \citealt{Kunth2000,Filho2013,Guseva2015}) when compared with dwarf galaxies of similar stellar masses ($\simeq10^8\Msun$, \citetalias{Sacchi2016}) that typically have $12+\logoh\sim7.9$.

Morphologically speaking, DDO 68 is classified as a dwarf irregular. At first glance, its unusual shape is characterized by the presence of a large, bright and distorted stellar component, the Cometary Tail (see the dashed red line in Fig.~\ref{fig:ddo68}), emerging from the south-eastern edge of the galaxy's main body. This component extends for $\simeq10\kpc$, with a stellar mass between $0.6\times10^7\Msun$ and $1.2\times10^7\Msun$ \citepalias{Sacchi2016}, at most $10\%$ of the mass in stars attributed to the whole system. While the Cometary Tail is somewhat more deficient in intermediate-age and old stellar populations compared to the main body, stars younger than 300 $\Myr$ are abundant in both components \citep{Makarov2017}, with an average star-formation rate peaking at a lookback time $\simeq30-50\Myr$, as derived by \citetalias{Sacchi2016}. Young stars also populate the roundish structure to the North of DDO 68, the so-called Cometary Head, hosting five $\HII$ regions, typically used to measure the galaxy metallicity (\citealt{Tikhonov2014}; \citetalias{Sacchi2016}; \citetalias{Annibali2019}). It is common to refer to the Tail or to the structure extending from the Tail to the Head as DDO 68 B, and to the remaining part of the galaxy as DDO 68 A or simply as DDO 68. Throughout this paper we will conform to this convention. A second low-luminosity arc (the Arc) emerges from the western edge of DDO 68 \citep[hereafter A16]{Annibali2016}. Differently from the bright and spatially extended Cometary Tail, the Arc is only visible using deep photometry, as demonstrated by \citetalias{Annibali2016} by means of Large Binocular Telescope (LBT) observations. The authors also report the presence of a second, low surface brightness stellar stream approximately $5\kpc$ to the north-east of DDO 68 (DDO 68 S1). As confirmed by \cite{Annibali2019b}, the stream has similar distance to DDO 68 and it is mostly populated by old stars ($>1-2\Gyr$), with a total mass in stars of $\sim10^6\Msun$ as derived from resolved stellar photometry. In the left panel of Fig.~\ref{fig:ddo68}, we show a LBT $g$ band image of DDO 68 from \citetalias{Annibali2016} where we have indicated all the aforementioned structures.

The morphological irregularities of DDO 68 strongly support the hypothesis that it has recently interacted with at least another system \citep[hereafter C14]{Cannon2014}. However, since DDO 68 falls in the periphery of one of the largest cosmic voids known, the Lynx-Cancer void ($D\simeq11\Mpc$, \citealt{Pustilnik2011}), the closest galaxy with similar radial velocity is UGC 5427 \citep{Makarova1998}, at a projected distance of $200\kpc$ \citepalias{Cannon2014}, too far to be an eligible candidate for a strong interaction. A safe and motivated way out to explain the peculiar structure and kinematics of DDO 68 is provided by \citetalias{Annibali2016} who, by means of $N$-body simulations, showed that its morphology is consistent with the outcome of an interaction with two independent smaller systems (with dynamical masses 1/10 and 1/150 times the dynamical mass of DDO 68), whose disrupted remnants end up forming the Tail (see also \citetalias{Cannon2014}) and the stream S1. As such, DDO 68 would be a fascinating example of dwarf galaxies caught during a multiple minor merger \citep{Tikhonov2014,Cannon2014,Annibali2016,Annibali2019}. \citetalias{Cannon2014} also reported the discovery of a gas rich companion ($\HI$ mass $\simeq 2.8\times10^7\Msun$), DDO 68 C, with similar systemic velocity to DDO 68, at a projected distance of $\simeq42\kpc$. Furthermore, the authors highlighted the presence of a low surface brightness $\HI$ bridge connecting DDO 68 C and DDO 68 suggestive of a more or less recent gravitational interaction \citepalias{Cannon2014} between the two systems.

We are left with a few open questions. As a matter of fact, DDO 68 has an oxygen abundance $\sim0.8$ dex lower than expected for its stellar mass \citep[see their Fig. 8]{McQuinn2020}: can this metal deficiency be linked to the merger hypothesis? It is generally believed that extremely low metallicities can be the result of dilution of primeval gas recently accreted; or that the metals have been lost due to galactic outflows powered by supernovae explosions and stellar winds \citep{Matteucci1983,Matteucci1985,Marconi1994,MacLow1999,Spitoni2015,Ceverino2016}. As for DDO 68, we also need to take into account that the evolution of galaxies within cosmic voids can be very different from that of their counterparts in denser environments (\citealt{Peebles2001,Hoeft2006}, but see also \citealt{McQuinn2020}). Void galaxies may have evolved more slowly and be metal poorer than galaxies in denser environments because of the absence of merger or gravitational interaction events able to trigger star-formation episodes. However, resolved stellar photometry with ACS observations has shown that DDO 68 has a well populated red giant branch \citepalias[RGB;][]{Sacchi2016}, indicating stellar populations at least as old as $1-2\Gyr$, a feature that excludes the hypothesis that this dwarf galaxy is a young stellar system hosting newly formed stars.

\begin{table*}
    \centering
    \caption{Main input parameters used to generate the ICs of the simulations. Galaxy: reference galaxy (DDO 68, Satellite B or Satellite S); $\Mdm$: total dark matter mass (equation~\ref{for:rhodm}); $a$: dark matter scale radius (equation~\ref{for:rhodm}); $\Mstar$: total stellar mass (equation~\ref{for:rhostar}); $\hstar$ and $\zstar$: stellar disc scale length and scale height, respectively (equation~\ref{for:rhostar}); $\Mgas$: gaseous disc total mass (equation~\ref{for:sigmagas}); $\hgas$: gaseous disc scale length (equation~\ref{for:sigmagas}). }\label{tab:ddo68ics}
    \begin{tabular}{cccccccc}
    \hline\hline
    \multirow{2}{*}{Galaxy} & $\Mdm$    &   $a$ &   $\Mstar$     &   $\hstar$ &    $\zstar$     & $\Mgas$   & $\hgas$  \\
     &  [$\Msun$] &  [$\kpc$] & [$\Msun$] & [$\kpc$] &  [$\kpc$] &  [$\Msun$] &  [$\kpc$] \\
    
    \hline\hline
    DDO 68 & $1.3\times10^{10}$ &   4   &   $1.17\times10^{8}$   &   1.02   &    0.153              &   $9.4\times10^8$ & 5.28   \\    
    Satellite B & $6.5\times10^8$ &   1.86   &   $1.2\times10^{7}$   &   0.47   &    0.0705         &   $0.6\times10^8$ & 1.41   \\    
    Satellite S & $8.6\times10^7$ &   1.2   &  $3.6\times10^6$   &   0.3   &   0.045  &   - & -  \\
    \hline\hline
    \end{tabular}
\end{table*}

Inspired by the results of the $N$-body models presented by \citetalias{Annibali2016}, in this paper we take a step further and provide hydrodynamical $N$-body simulations of DDO 68 aimed at reproducing most of the available structural and kinematical observations of DDO 68 derived in the past years (\citetalias{Cannon2014}; \citetalias{Annibali2016}; \citetalias{Sacchi2016}; \citetalias{Annibali2019}; \citealt{Annibali2019b}). In fact, even if \citetalias{Annibali2016} provided a satisfying physical justification to the hypothesis that DDO 68 has been shaped by multiple merger events, their simulations do not consider some fundamental aspects. In this respect, we provide a higher degree of completeness and complexity by accounting for gas physics, and thus trying to reproduce the gas kinematics. We explore a larger portion of the vast parameter space defined by the orbital parameters using realistic galaxy models. We account for the self-gravity of the stellar and gas components of all the galaxies at play, we consider the possibility that a further interaction with DDO 68 C may justify some of the peculiarities observed in DDO 68, and we explore to what extent the merger between DDO 68 and a pristine gas-rich galaxy can contribute to explain the metal deficiency of DDO 68.

The paper is structured as follows. In Section~\ref{sec:setup} we describe the set-up of the simulations and the criteria used to set the galaxies' initial conditions (ICs) and orbital parameters, while in Section~\ref{sec:ddo68c} we argue that the gravitational interaction with DDO 68 C is not enough to explain the irregular structure of DDO 68. The simulations provide configurations of DDO 68 in good agreement with the observations, as discussed in Section~\ref{sec:res}. In Section~\ref{sec:met} we show how the interaction with a metal-poor satellite can account for the metal deficiency of DDO 68. Section~\ref{sec:conc} summarises our conclusions.

\section{Set-up of the simulations}
\label{sec:setup}

Our simulations follow the late evolution of stellar systems whose stellar populations are already in place (no star formation included). The simulations are comprised of three systems: a dominant DDO 68-like galaxy in interaction with two smaller satellites, which we will refer to as Satellite B and Satellite S. As in \citetalias{Annibali2016}, while the most massive satellite (Satellite B) is intended to reproduce DDO 68 B, i.e. the Cometary Tail to the south-west of DDO 68, the least massive satellite (Satellite S) will reproduce the faint stellar stream DDO 68-S1 observed to the north-east of DDO 68. 

In the following, we describe the procedure used to set and run our hydrodynamical $N$-body simulations. In Section~\ref{sec:models} we briefly list the main properties of the analytic galaxy model used to shape the three systems, describing the most relevant free parameters the models depend on, and the scheme adopted to generate the initial phase-space positions of all components. In Section~\ref{sec:freeparam} we describe how we fixed the free parameters of DDO 68, Satellite B and Satellite S and comment on the kinematic and structural properties resulting from the isolated evolution of each galaxy. In Section~\ref{sec:xovo}, we describe the set of simulations and the procedure used to fix the orbital parameters of the satellites.

\subsection{Properties of the analytic galaxy models}
\label{sec:models}

The three systems are modeled as disc galaxies composed of a dark matter halo, a gaseous and a stellar disc. The dark matter halo follows the \cite{Dehnen1993} profile
\begin{equation}\label{for:rhodm}
    \rhodm(r) = \frac{3\Mdm}{4\pi}\frac{a}{(r+a)^4},
\end{equation}
where $\Mdm$ is the total dark matter mass, and $a$ is the halo scale radius. 

Both star and gas discs have non-negligible thickness. The stellar disc stratifies with radially constant scale height $\zstar$ and its full density distribution is
\begin{equation}\label{for:rhostar}
    \rhostar(R,z) = \frac{\Mstar}{4\pi\hstar^2\zstar}e^{-\frac{R}{\hstar}}\sech^2\biggl(\frac{z}{\zstar}\biggr),
\end{equation}
where $\Mstar$ is the total stellar mass, and $\hstar$ and $\zstar$ are the stellar disc scale length and height, respectively. The gas surface density follows the relation
\begin{equation}\label{for:sigmagas}
    \Sigmagas(R) = \frac{\Mgas}{2\pi\hgas^2}e^{-\frac{R}{\hgas}},
\end{equation}
where $\Mgas$ and $\hgas$ are the gas total mass and the gas disc scale length, respectively. The vertical stratification of the gas is fixed by the hydrostatic equilibrium
\begin{equation}\label{for:verteq}
    \frac{\DD \Phitot}{\DD z} = -\frac{1}{\rhogas}\frac{\DD P}{\DD z}
\end{equation}
requiring that
\begin{equation}\label{for:rhogas}
    \Sigmagas(R) = \int_{-\infty}^{+\infty}\rhogas(R,z)\,\dd z.
\end{equation}
In equation (\ref{for:verteq}), $P$ is the pressure of the gas, which we have assumed, at each radius $R$, to be isothermal with $T/\mu=3400\K$, where $T$ is the temperature and $\mu$ is the mean molecular weight. $\Phitot$ is the total gravitational potential (dark matter, stars and gas), found iteratively as in \cite{Springel2005}. As in \cite{Pascale2021}, the halo particles are drawn from the dark-matter isotropic distribution function (DF), assuming $\Phitot=\Phidm$ (i.e. neglecting the contribution to the total potential from stars and gas). The stellar disc DF depends on the specific (i.e. per unit mass) energy $E$ and on the vertical component of the specific angular momentum $\Lz$ only. Disc particles are drawn in Maxwellian approximation, with velocity dispersion elements derived from the Jeans equations, and relying on the epicyclic approximation to sample the azimuthal component of the particle velocities. The gas velocity field is only made by the azimuthal component, found imposing the stationary Euler equation \citep[for details see][]{Springel2005}.

The gas metallicity distribution varies as a function of the galactocentric distance and it evolves in time as a passive scalar field, i.e. it does not have any effect on the dynamical and structural properties of the galaxies in the simulations since we do not account for any source of heating/cooling such as star formation, supernovae feedback or radiative cooling which, in general, depend on the chemical composition of the gas. A complete (non-adiabatic) treatment of gas physics would inevitably imply that simulations with different chemical set ups are not dynamically equivalent. We will separately discuss in detail the metallicity set-up, its numerical implementation and the results in Section~\ref{sec:met}.

\subsection{Setting the galaxy model free parameters}
\label{sec:freeparam}

Apart from the metallicity distribution, a single galaxy model is uniquely determined by the following set of free parameters
\begin{equation}\label{for:params}
 \bbxi = \{\Mdm, a, \Mstar, \hstar, \zstar, \Mgas, \hgas\},
\end{equation}
for a total of 21 free parameters for DDO 68, Satellite B and S (see Table~\ref{tab:ddo68ics}).

\subsubsection{DDO 68}
\label{sec:ddo68}
The parameters of the dark matter halo of DDO 68 follow from \citetalias{Annibali2016}: $\Mdm=1.3\times10^{10}\Msun$ and $a=4\kpc$. Such values ensure that $\Mdm(11\kpc)\simeq5.3\times10^9\Msun$, as measured by \citetalias{Cannon2014}, where $\Mdm(11\kpc)$ is the dark matter mass within $11\kpc$. We note that the estimate of \citetalias{Cannon2014} refers to the galaxy dynamical mass, and it only provides a lower limit to it since they do not apply corrections for non-circular motions. A different estimate comes from \cite{Ekta2008} who, from the $\HI$ velocity field map, provide a rotation velocity peaking at $55\kms$ and $48\kms$ for galactocentric distances $R=144\asec$ and $R=168\asec$, respectively. At $d=12.65\Mpc$, considering an average $R=156\asec=9.57\kpc$, this implies a dark-matter mass of $\simeq5.9\times10^9\Msun$, slightly larger than the value adopted here.

According to the detailed star-formation history derived by \citetalias{Sacchi2016}, the stellar mass of DDO 68 B ranges between $0.6\times10^7\Msun$ and $1.2\times10^7\Msun$, depending on whether DDO 68 B is only composed by the Tail or it extends from the Tail to the Head. Since in the simulations we expect part of its initial stellar mass to be stripped or to be lost below the surface brightness limit when producing mock LBT images (e.g. see the right panel of Fig.~\ref{fig:ddo68}), we attributed to Satellite B a larger initial stellar mass of $1.2\times10^7\Msun$. The remaining stellar mass of $\Mstar=1.17\times10^8\Msun$, approximately 91\% of the total stellar mass inferred by \citetalias{Sacchi2016}, is attributed to DDO 68. We adopt a stellar disc scale length $\hstar=1.02\kpc$, as in \citetalias{Annibali2016}, while $\zstar=0.15\hstar=0.153\kpc$ \citep{Kregel2002,Oh2015}. The $\HI$ total mass derived by \citetalias{Cannon2014} is $1.0\pm0.15\times10^9\Msun$. This value is partitioned between DDO 68 and Satellite B: we attribute $\Mgas=9.4\times10^8\Msun$ to the former, the remaining gas to the latter. We fixed $\hgas=5.15\hstar=5.28\kpc$, which ensures that the ratio between the stellar and gas mass within the stellar disc scale length is $\simeq0.5$. With this choice the stars dominate within $\hstar$, while the gas dominates further out, as measured in classical disc galaxies. 

\subsubsection{Satellite B and Satellite S}
\label{sec:satrsatp}
The dark matter halo of Satellite B has $\Mdm=6.5\times10^8\Msun$ and $a=1.86\kpc$, as in \citetalias{Annibali2016}. As pointed out in the previous Section, the initial total stellar mass of Satellite B is $\Mstar=1.2\times10^7\Msun$, while $\hstar=0.47\kpc$, as in \citetalias{Annibali2016}. Also for Satellite B, $\zstar=0.15\hstar=0.0705\kpc$. The $\HI$ disc has $\Mgas=6\times10^7\Msun$ and $\hgas=3\hstar=1.41\kpc$. We choose a gas partitioning  between DDO 68 and DDO 68 B that ensures $\Mgas/\Mstar=5$ for Satellite B, a value in agreement with gas-to-stellar mass ratios observed in typical dwarf galaxies with Satellite B stellar mass \citep{Maddox2015,Parkash2018}.

Satellite S is a gas-free, scaled-down version of DDO 68, with a dynamical mass approximately 150 times smaller than DDO 68. As in \citetalias{Annibali2016}, $\Mdm=8.6\times10^7\Msun$ and $a=1.2\kpc$. We set its initial stellar mass to $\Mstar=3.6\times10^6\Msun$, while $a=0.3\kpc$, as in \citetalias{Annibali2016}. The total stellar mass measured by \cite{Annibali2019b} is approximately $1.2\times10^6\Msun$ but, given its large g-bands mass-to-light ratio (see Appendix~\ref{app:A}), and given that we expect a large mass loss, we set its initial mass to at least three times the measured, present day mass of DDO 68-S1. While the position of DDO 68 in the stellar-to-halo mass relation is marginally consistent with that expected from galaxies with similar stellar masses (see, for instance, \citealt{Read2017}), the halo mass of Satellite B and Satellite S is lower than the one of galaxies with similar stellar masses. As we will discuss in Section~\ref{sec:xovo}, since in the simulations we set the two satellites on a relatively close orbit around DDO 68, we may presume their dark matter halo to be less massive than estimated due to prior tidal interactions with the host galaxy.

\begin{figure*}
    \centering
    \includegraphics[width=.9\hsize]{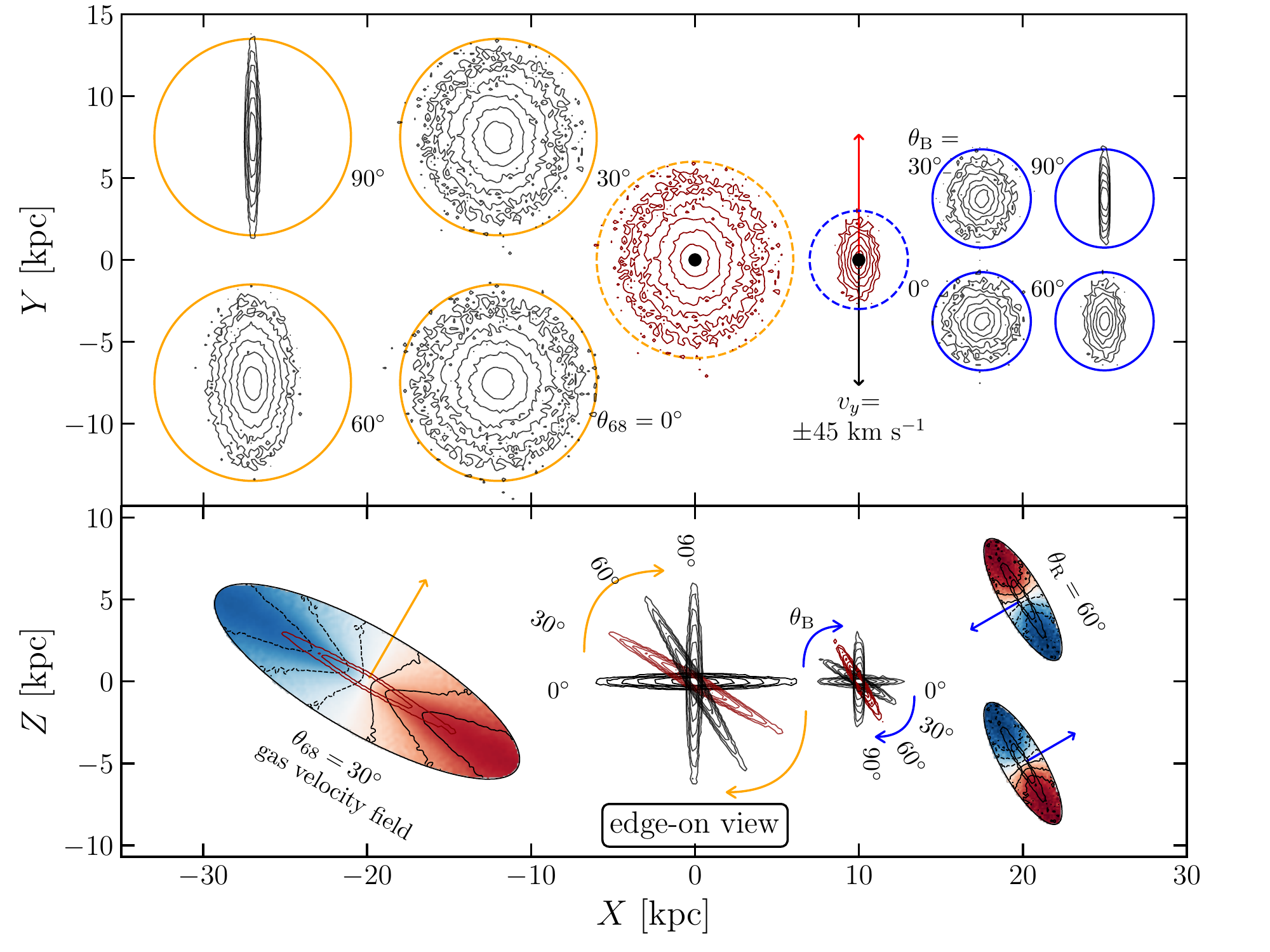}
    \caption{Schematic view of the ICs for the set of simulations of Section~\ref{sec:xovo}. The top panel shows the stellar density distributions of DDO 68 and Satellite B obtained assuming as line-of-sight the $z$-axis ($X\equiv x$ and $Y\equiv y$). The center of mass of DDO 68 is at $\xyz=(0,0,0)$, while the center of mass of Satellite B (black dots) is at $\xyz=(10\kpc,0,0)$. The solid orange and the solid blue circles contain the galaxies projections along the $z$-axis for $\thetaddo, \thetaR= 0^{\circ},30^{\circ},60^{\circ},90^{\circ}$ that should be placed within the dashed orange and blue circles to reproduce the ICs of a single simulation. Satellite B has initial velocity $\vxyz=(0,\pm45\kms,0)$ (red and black arrows). Bottom panel: edge-on projections (assuming as line-of-sight the $y$-axis) of the galaxies as in the top panel (so $Z\equiv z$). The red-isodensity contours in the top and bottom panels and the red arrow in the top panel show the galaxies ICs of the simulation that produce a configuration of DDO 68 as in the observations ($\thetaddo=30^{\circ}$, $\thetaR=60^{\circ}$; see Section~\ref{sec:res}).
    In the bottom panel we show the gaseous initial velocity field for a reference simulation with $\thetaddo=30^{\circ}$ and $\thetaR=60^{\circ}$: blue colors mark the disc approaching arm, red colors the disc receding arm. The orange and blue straight arrows show for DDO 68 and Satellite B, respectively, the direction of the discs angular momentum.}\label{fig:ics}
\end{figure*}

\begin{table*}
    \centering
    \caption{Main parameters used to run the simulations. Galaxy: reference galaxy (DDO 68, Satellite B or Satellite S); $\Ndm$, $\Nst$, $\Ngas$: total number of dark-matter, stellar and gas particles. $\sldm$, $\slst$, $\slgas$: softening lengths of dark matter, stars and gas.}\label{tab:ddo68sim}
    \begin{tabular}{cccccccc}
    \hline\hline
        Galaxy &   $\Ndm$  &   $\sldm$ [$\kpc$] &    $\Nst$  &   \multicolumn{2}{c}{$\slst$ [$\kpc$]} &   $\Ngas$ &   $\slgas$ [$\kpc$] \\ 
    \hline\hline
    DDO 68      &   9736000 &  0.184 &   93600   &   \multicolumn{2}{c}{0.04}         &   752000  &  0.092   \\
    Satellite B &   486800  &  0.184 &   9600    &  \multicolumn{2}{c}{0.04}         &   48000  &  0.092   \\
    Satellite S  &  128820  &  0.185    &  5760 &   \multicolumn{2}{c}{0.05} &  - & - \\
    \hline\hline
    \end{tabular}
\end{table*}

\subsubsection{Specifics of the simulations and equilibrium configurations} 
\label{sec:eq}

For each component of each galaxy, the gravitational softening is computed requiring that the maximum force between the particles of that component should not be larger than the mean-field strength they generate \citep{Dehnen2011}. The number of particles of all the collisionless components of DDO 68 and Satellite B is set requiring that all particles have the same mass, $\mpart=1250\Msun$. For Satellite S, the least massive galaxy, we require the dark and visible particles to have $\mpart=625\Msun$; this allows us to sample its dark matter halo and stellar disc with a sufficiently large number of particles. All dark matter haloes are sampled out to $45$ scale radii, the distance within which the haloes contain $\simeq94$\% of their total mass (i.e. $\Ndm\mpart=0.94\Mdm$, with $\Ndm$ the number of dark matter particles of each galaxy).

To run the simulations, we rely on the publicly available version of the code {\sc arepo} \citep{Springel2010,Weinberger2020}, a hydrodynamical $N$-body code that exploits both Lagrangian and Eulerian hydrodynamics by solving Euler equations on an unstructured Voronoi mesh that is free to move with the fluid flow. We sample the gaseous discs of DDO 68 and Satellite B with particles of $\mpartgas=1250\Msun$, and use a mesh refinement criterion that ensures that the gas cells keep approximately the same mass. With this choice, we can sample high-density regions with a relatively large number of particles. Table~\ref{tab:ddo68ics} summarizes the main parameters used to generate the ICs of the three galaxies, while Table~\ref{tab:ddo68sim} lists the softening lengths and number of particles adopted in the simulations.

Since: i) the dark-matter halo particles of all galaxies have been sampled as if the halos were in isolation and ii) the stellar disc particles are sampled in Maxwellian approximation, we expect all the galaxies to be in quasi-equilibrium. We let each galaxy evolve in isolation for $\tmax\simeq3\Gyr$ to make it reach an equilibrium configuration. In these simulations we use individual time steps for each particle partitioned on a power-of-two hierarchy based on local stability criteria \citep{Weinberger2020}, with typical time step values of $0.1\Myr$.

In the non-interacting configuration, the dark matter halo of the three galaxies slightly collapses in its central regions, where the stellar and gaseous discs contribute the most to the gravitational potential. The effect is more pronounced in DDO 68, where the mass contribution of the discs to the total mass is about 40\% within the stellar scale length. As soon as the halo collapses and contracts, the system moves towards equilibrium through violent relaxation \citep{LyndenBell1967}, and the discs respond to the perturbed total potential developing density waves that propagate inside-out. In all the three simulations for the galaxies evolving in isolation, the perturbations in the star and gas discs fade away after at most $\simeq3\Gyr$. To avoid spurious effects in our conclusions of gaseous and
stellar features resulting from the three galaxies' interaction, we adopt as ICs the outcome of the simulations for the systems evolving in isolation after $3\Gyr$.

\begin{table*}
    \centering
    \caption{Initial position ($\xini,\yini,\zini$) and velocity ($\vxini,\vyini,\vzini$) of Satellite B and Satellite S in the simulation shown in Fig.s~\ref{fig:ddo68}, \ref{fig:incl}, ~\ref{fig:orbitmodel} and Fig.~\ref{fig:vfield}. The simulation corresponds to $\thetaddo=30^{\circ}$ and $\thetaR=60^{\circ}$. The disc of Satellite S has an initial inclination given by counterclockwise rotations about the $x$, $y$, and $z$-axis of $-10^{\circ}$, $45^{\circ}$ and $45^{\circ}$, respectively, performed in this order. The center of mass of DDO 68 is at $\xyz=(0,0,0)$.}\label{tab:icsbestsim}
    \begin{tabular}{ccccccc}
    \hline\hline
    Galaxy & $\xini$ [$\kpc$] & $\yini$ [$\kpc$] & $\zini$ [$\kpc$] & $\vxini$ [$\kms$] & $\vyini$ [$\kms$] & $\vzini$ [$\kms$]  \\
    \hline\hline
    Satellite B     &   10    &   0     &     0     &   0       &    -45    &  0      \\
    Satellite S     &  4.155 &   -9.024  &   5.793  &   41.17  &   8.733   &  -15.93   \\
    \hline\hline
    \end{tabular}
\end{table*}

\subsection{Putting the galaxies together}
\label{sec:xovo}

\subsubsection{DDO 68 and Satellite B}

Given the system complexity, we adopt a double-step procedure: i) we first run simulations considering only DDO 68 and Satellite B and reproduce the overall optical structure and the gas kinematics of DDO 68 and DDO 68 B exploring a specific portion of the parameter space; ii) once we have found a configuration that resembles the observed one, we add Satellite S to the simulation such as to reproduce the stream S1. 

The center of mass of DDO 68 is at $\xyzi=(0,0,0)$, with $\xyz$ a Cartesian reference frame whose $y$-axis belongs to the discs plane of DDO 68. In the same reference frame, the center of mass of Satellite B is at $\xyzi=(10\kpc,0,0)$. Called $\Mi$ the dynamical mass of the primary object (DDO 68) within $d=10\kpc$, $\Mii$ the dynamical mass of the secondary object (Satellite B, see Table~\ref{tab:ddo68sim}), the truncation radius \citep{BinneyTremaine2008} of Satellite B at $d$ is $\rt\equiv d(\Mii/\Mi)^{1/3}\simeq5.1\kpc$, approximately $\sim3$ times the scale radius of its dark matter halo, $\sim3.5$ times its gaseous disc scale length, and $\sim11$ times its stellar disc scale length. The system expected truncation radius is large enough to make our results independent of the initial position of Satellite B. Satellite B can orbit around DDO 68 with initial velocities $\vxyzi=(0,45\kms,0)$ or $\vxyzi=(0,-45\kms,0)$. The initial velocity of the center of mass of Satellite B is that of a particle in an almost circular orbit, because $\vc(10\kpc)=45\kms$, with $\vc$ the circular velocity of DDO 68. The discs of DDO 68 and Satellite B can be inclined by $\thetaddo$ and $\thetaR$, respectively: the rotation is clockwise about the $y$-axis, with $\thetaddo$ and $\thetaR$ measured from the positive $z$-axis. In our convention, when, for instance, $\thetaddo=0^{\circ}$, the galaxy's symmetry axis is aligned with the $z$-axis. We perform a systematic search in the portion of the parameter space defined by ($\thetaddo,\thetaR,\vyini$), considering all the combinations given by $\thetaddo\in\{0^{\circ};30^{\circ};60^{\circ};90^{\circ}\}$, $\thetaR\in\{0^{\circ};30^{\circ};60^{\circ};90^{\circ}\}$ and $\vyini=\pm45\kms$. Also, for each $\vyini$, $\thetaddo$ and $\thetaR$, we run two simulations where the discs of Satellite B rotate in the two possible senses of rotation about its symmetry axis, for a total of 64 simulations. Figure~\ref{fig:ics} is a schematic representation of the ICs: for all the combinations of $\thetaddo$ and $\thetaR$ we show how the stellar discs of DDO 68 and Satellite B appear when the line-of-sight is along the system $z$-axis (top panels) and $y$-axis (bottom panels). We refer the reader to Appendix~\ref{app:B}, where we show that the very same simulation of \citetalias{Annibali2016} with the introduction of hydrodynamics and more complicated galaxy models with respect their original $N$-body simulations does not reproduce anymore the structure and kinematics of DDO 68.

We expect Satellite B to spiral towards DDO 68 due to orbital decay caused by dynamical friction, possibly forming the Cometary Tail while being stripped by the tidal force field of DDO 68. All the simulations have been evolved for $2.2\Gyr$, a value compatible with several dynamical friction timescales of Satellite B. We estimate a dynamical friction timescale \citep{BinneyTremaine2008}
\begin{equation}\label{for:tdf}
    \tdf = \frac{1.17}{\ln\Lambda}\frac{\Mi}{\Mii}\tcross = 0.3-0.6\Gyr,
\end{equation}
where $\tcross=0.43\Gyr$ is the crossing time (i.e. the time needed by Satellite B to cross twice its initial distance, $20\kpc$, with a speed of $45\kms$) and the lower and upper limits on $\tdf$ depend on the values of the Coulomb logarithm used, $\ln\Lambda=15$ and $\ln\Lambda=6$, respectively. As for the simulations of Section~\ref{sec:eq}, we use an adaptive timestep refinement hierarchy with typical timestep values $\simeq0.1\Myr$.

\subsubsection{Satellite S}

\begin{figure*}
    \centering
    \includegraphics[width=1\hsize]{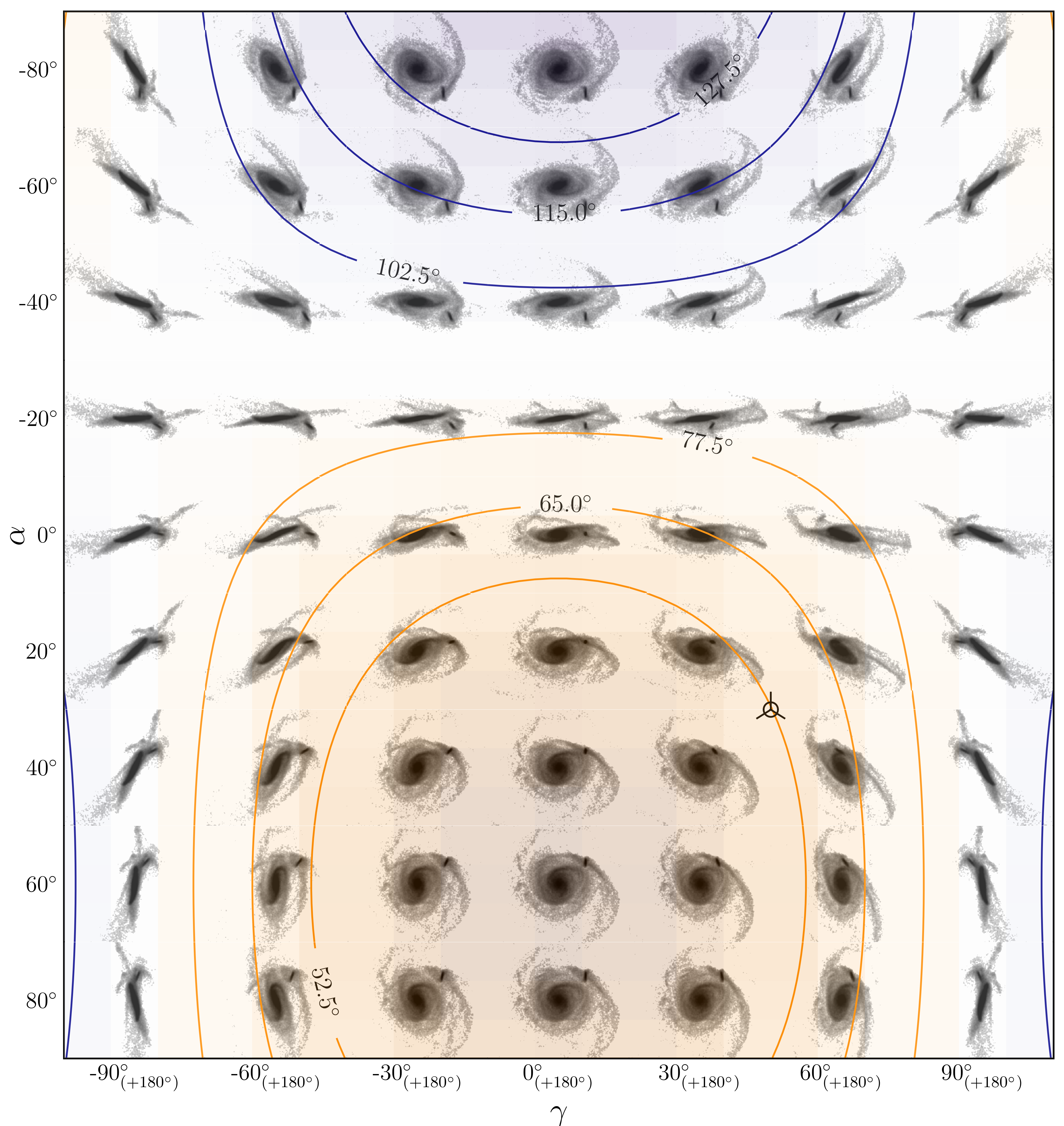}
    \caption{Stellar surface brightness maps obtained for lines-of-sight given by several combinations of $\alpha$ and $\gamma$, as labeled along the main axes, given a simulation evolved for $t=1\Gyr$, with $\thetaddo=30^{\circ}$ and $\thetaR=60^{\circ}$ (see Section~\ref{sec:mockddo}). Orange and blue colors in the background correspond to face-on inclinations $i$, light colors to edge-on inclinations, according to the relation $\cos i= \sin(\thetaddo+\alpha)\cos\gamma$. Note that $i\in[0,180^{\circ}]$, so face-on views are obtained both for $i=0^{\circ}$ (orange) and $180^{\circ}$ (blue). The blue and orange contours mark then the two possible galaxy's iso-inclinations $i=65^{\circ}\pm12.5^{\circ}$ and $i=(180^{\circ}-65^{\circ})\pm12.5^{\circ}$, as in \citetalias{Cannon2014}. The surface brightness maps have spatial resolution given by pixel size of $1\asec\times1\asec$, and they have been convolved with a Gaussian beam of $2.82\asec$. The mass-light-ratios used are $\MLgA=\MLgB=0.5$ (see Appendix~\ref{app:A}). Once a line-of-sight is fixed, an increase of $\gamma$ of $180^{\circ}$ produces a mirror configuration (i.e. the spatial distribution and the velocity field are flipped). The marker at $(\alpha,\gamma)=(30^{\circ},45^{\circ}, 180^{\circ})$ shows the line-of-sight of the model analysed in Section~\ref{sec:mockddo}. Note that a further rotation of $180^{\circ}$ about $\gamma=45^{\circ}$ is required to flip the image as in the observations.}\label{fig:incl}
\end{figure*}

The specific initial position and the inclination of the disc of Satellite S, which is supposed to form the stellar stream DDO 68-S1, depend on the line-of-sight that produces, after the simulation has evolved for a time $t$, a configuration of DDO 68 and Satellite B similar to the observed one. 

As described in detail in Section~\ref{sec:simobs}, we have visually inspected all possible lines-of-sight throughout the time evolution of the 64 simulations with only DDO 68 and Satellite B in search for configurations with stellar surface brightness, gas distribution and velocity field similar to the observed ones. Once we find one, we re-run the simulation adding Satellite S to produce, on the plane of the sky after a time $t$ and along the same line-of-sight, the stellar stream S1 located $\simeq5\kpc$ to the north-east of DDO 68, similarly to Fig.~\ref{fig:ddo68}. Satellite S starts from $\rini\equiv||\vecri||=\sqrt{\xini^2+\yini^2+\zini^2}=11.5\kpc$, with speed $\vini\equiv||\vecvi||=\sqrt{\vxini^2+\vyini^2+\vzini^2}=45\kms$, on an orbit that would be almost circular for a particle at $11.5\kpc$ since $\vc(11.5\kpc)=45\kms$ ($\vecri\cdot \vecvi=0$, with $\cdot$ indicating the standard inner product between vectors).

Notice that being Satellite S approximately at the same distance as Satellite B and 7.5 times less massive than Satellite B, its dynamical friction timescale is at least 7.5 times larger, so we expect its orbit to decay slowly with time (see equation~\ref{for:tdf}).

\section{The case of DDO 68 C}
\label{sec:ddo68c}

$42\kpc$ in projection to the north of DDO 68 and with approximately the same systemic velocity, \citetalias{Cannon2014} reported the discovery of a previously unknown gas dominated galaxy, DDO 68 C ($M_{\HI}=2.8\times10^7\Msun$). An ongoing or recent gravitational interaction with DDO 68 C has been called upon to explain the irregular shape of DDO 68 \citepalias{Cannon2014}, motivated by the presence of a low surface brightness $\HI$ bridge connecting the two objects. Here we discuss critically this hypothesis.

As argued by \citetalias{Annibali2016}, given its present-day location, DDO 68 C should be at least 35 times as massive as DDO 68 in order to provide a gravitational torque on DDO 68 able to explain, for instance, the presence of the Cometary Tail at $\sim5\kpc$. This is extremely unlikely given its size and baryonic content. An independent constraint on the dynamical mass of DDO 68 C is given by its $\HI$ rotation velocity \citepalias{Cannon2014}. The far-UV morphology revealed by Galex suggests a very low (although unknown) inclination which, if as low as $i=20^{\circ}$, would imply a dynamical mass $r\vrot^2/G$ within $1\kpc$\footnote{$1\kpc$ is about the spatial extent of the detected $\HI$ gas.} of $10^8\Msun$, given a rotation speed peaking at $\vrot\simeq7-10\kms$. In this respect, Satellite S provides a proxy for the effects that a satellite galaxy as massive as $10^8\Msun$ would have on DDO 68 when moving close to it. However, as we will discuss in the following Section, when comparing simulations differing only for the presence of Satellite S, we notice that its motion around DDO 68 does not provide any detectable or significant perturbation. 

If the cause of the recent star-formation activity of DDO 68 were an interaction with DDO 68 C, the component of its velocity on the plane on the sky should be at least as high as $400-800\kms$ to cross $42\kpc$ in $100-50\Myr$ (\citetalias{Sacchi2016}; \citealt{Cignoni2019}), a value much more consistent with the high density environments of groups or galaxy clusters, where the relative velocity between galaxies can be as high as the cluster velocity dispersion (e.g. $\sigma\sim1000\kms$), than with cosmic voids, such as that in which DDO 68 is hosted. The very presence of gas in DDO 68 C suggests that the putative fly-by must have happened at a large pericentric distance or with low relative velocity to prevent the complete removal of its gas.

Putting all these considerations together we conclude that, even if it is reasonable to speculate that DDO 68 C has interacted in the past with DDO 68, as suggested by the presence of the $\HI$ bridge, this interaction: i) is not the cause of the recent star-formation activity; ii) has not caused the tailed shape of DDO 68; iii) has not caused the formation of the Arc or the S1 stellar stream. It seems reasonable to believe that most of the gravitational disturbance has been felt by DDO 68 C, rather than by DDO 68. Considering the negligible impact that DDO 68 C must have had on DDO 68, we decided to not include it in our hydrodynamical $N$-body models.

\section{Results}
\label{sec:res}

The observations provide constraints for: i) the galaxy inclination; ii) the sense of rotation of the discs of DDO 68; iii) the shape of the Cometary Tail; iv) the line-of-sight velocity gradient observed throughout the Tail and DDO 68, and v) stellar mass, position and size of the stellar stream S1. We look for simulations that reproduce all these features at once. We do not expect to provide unique solutions for the observed properties of DDO 68’s system given the that we only explored a limited portion of the vast parameter space. As mentioned in Section \ref{sec:xovo}, we have, indeed, only considered specific orbital histories: Satellite B orbits around DDO 68 onto circular orbits, while it is not excluded that more radial interactions may result in similar outcomes. Also, although observationally motivated,  we have fixed all the parameters of the different galaxy models (see equation~\ref{for:params}): specifically we did not explore mergers / interactions with different mass ratios. Nevertheless, we believe we have significantly constrained the range of viable scenarios.


\subsection{Preliminary considerations}
\label{sec:simobs}

\begin{figure*}
    \centering
    \includegraphics[width=1\hsize]{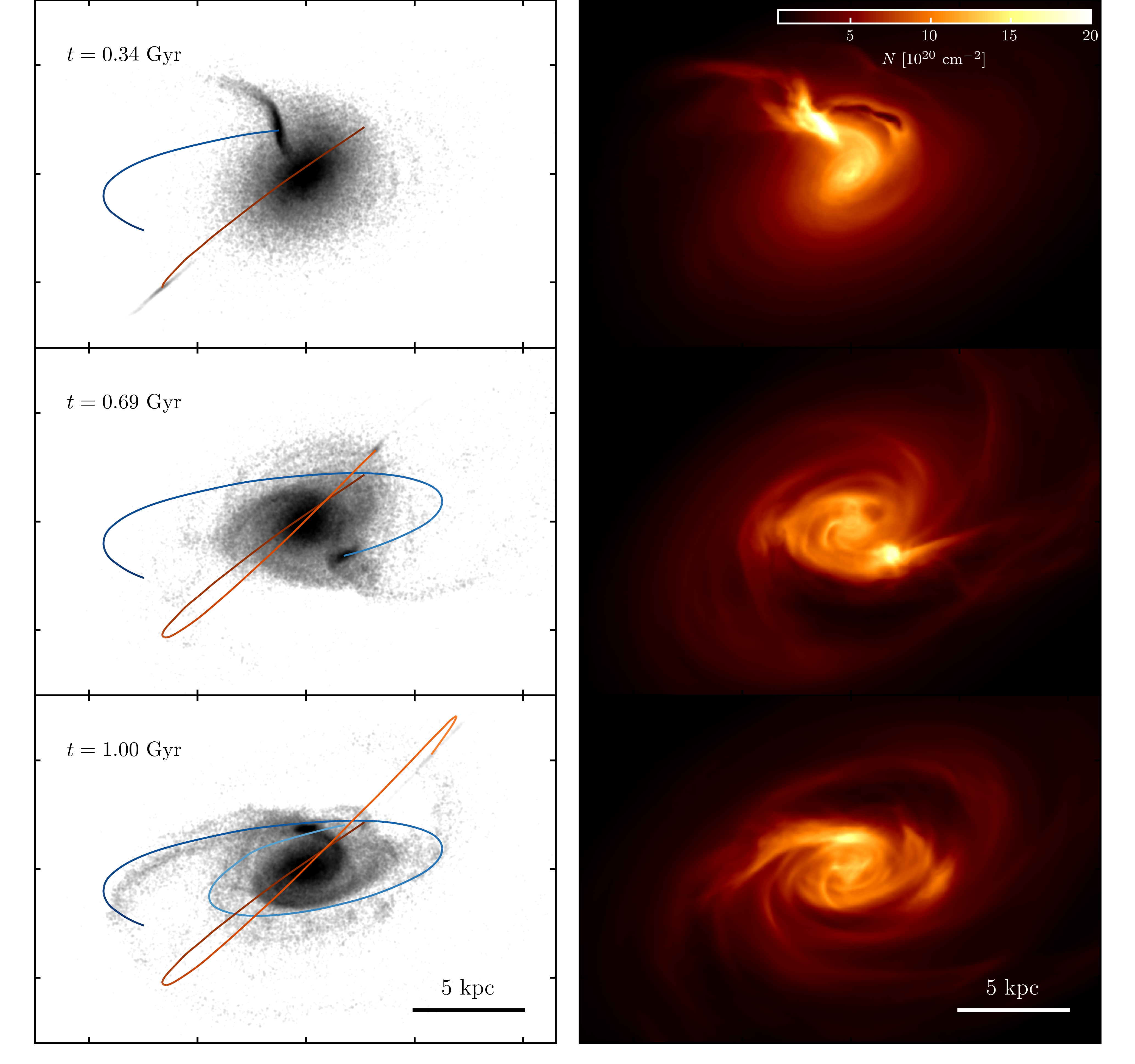}
    \caption{Left panels: stellar surface brightness maps from subsequent snapshots of the reference simulation of Section~\ref{sec:res}. In each panel the line-of-sight is given by $(\alpha,\gamma,\beta)=(30^{\circ},225^{\circ},-20^{\circ})$, the pixel size is $0.225\asec\times0.225\asec$, and the maps have been smoothed with a gaussian beam of FWHM $1.9\asec$, corresponding to a $0.7\asec$ seeing. The mass-to-light ratios attributed to the particles of DDO 68, Satellite B and Satellite S are $\MLgA=0.51$, $\MLgB=0.51$ and $\MLgS=5.5$, respectively, and the maps have depth of 29 mag$\asec^{-2}$. The field in each panel is $18\times24\kpc^2$ and the sequence of panels (from top to bottom) evolves showing snapshots corresponding to $t=0.34\Gyr$, $t=0.69\Gyr$ and $t=1\Gyr$. The blue and orange lines show the orbits of the centres of mass of the dark matter haloes of Satellite B and Satellite S, respectively, computed with the shrinking sphere method \citep{Power2003}. The bottom left panel shows the same configuration as in the right panel of Fig.~\ref{fig:ddo68}. Right panels: same as the left panels, but showing the gas projected number density. In this case, the pixel size is $0.68\asec\times0.68\asec$ and the maps are convolved with a $1.9\asec$ beam. Details on the ICs are given in Section~\ref{sec:xovo}, and Tables~\ref{tab:ddo68ics} and ~\ref{tab:icsbestsim}.}\label{fig:orbitmodel}
\end{figure*}

For clarity, we parametrize a line-of-sight with the triplet of Euler angles $\alpha$, $\gamma$ and $\beta$. Starting from the same Cartesian reference frame as in the ICs, we first align the line-of-sight with the positive $x$-axis, then we rotate the system counterclockwise by an angle $\alpha$ about the $y$-axis. A counterclockwise rotation by an angle $\gamma$ about the $z$-axis and a counterclockwise rotation by an angle $\beta$ about the $x$-axis are then applied\footnote{Note that rotations do not commutate in $\mathbb{R}^{3}$, so they must be performed in this order to get the same outcome.}. Once the rotations have been performed, we project the system along the resulting line-of-sight.

Using archival VLA observations of DDO 68, \citetalias{Cannon2014} roughly constrained the galaxy inclination to $i=65.2^{\circ}$ by means of tilted-ring model fitting of their derived $\HI$ velocity field map. In our formalism, the inclination $i$ fixes the angles $\alpha$ and $\gamma$ according to the relation $\cos i=\sin(\thetaddo+\alpha)\cos\gamma$, of which we give a representation in Fig.~\ref{fig:incl} in case of $\thetaddo=30^{\circ}$. We use light orange and blue backgrounds for face-on inclinations, and white backgrounds for edge-on inclinations. The $i=65.2^{\circ}$ inclination of \citetalias{Cannon2014} reduces the allowed lines-of-sight: since the authors do not provide uncertainties on $i$, and assumed that $i$ can slightly change through time due to the interaction with Satellite B, we considered all the pairs ($\alpha,\gamma$) that produce, at the beginning of the simulations, inclinations consistent with the wider range $i=65^{\circ}\pm12.5^{\circ}$, also marked by orange and blue contours in Fig.~\ref{fig:incl}. For any combination of $\alpha$ and $\gamma$ labeled along the main axes, we superimpose the stellar surface brightness map (DDO 68 and Satellite B) obtained projecting the system along the resulting lines-of-sight considering a simulation with $\thetaddo=30^{\circ}$ and $\thetaR=60^{\circ}$ that has evolved for $t=1\Gyr$. We assumed the mass-to-light ratios $\MLgA=\MLgB=0.51$ (see Appendix~\ref{app:A}) and the same surface brightness limit as in Fig.~\ref{fig:ddo68}. Note that to highlight small and low-luminosity features in the stellar density maps we used a smaller resolution than that adopted in the right-hand panel of Fig.~\ref{fig:ddo68}: the maps have a reference pixel size of $1\asec\times1\asec$ and they are convolved with a Gaussian beam of $2.82\asec$ FWHM. Since the galaxy inclination $i$ does not depend on the angle $\beta$, we set $\beta=0^{\circ}$ for simplicity. Note that $\beta$ can be fixed by the position angle (PA) since it merely reduces to a rotation of the system on the plane of the sky. Since the system is significantly not axisymmetric due to Satellite B, lines-of-sight with same $\alpha$ but opposite $\gamma$ (and vice-versa) are not equivalent. A change of sign in $\gamma$ produces not just a flip of the velocity field that turns the approaching arm into the receding arm, but also a different projected spatial distribution of the visible component. Note also that, once a line-of-sight is fixed, increasing $\gamma$ by $180^{\circ}$, for fixed $\alpha$, flips the stellar and gas spatial distributions and the corresponding velocity field.
 
In our analysis, we have visually inspected the stellar surface brightness maps, gaseous surface density and line-of-sight velocity fields throughout the time evolution of each simulation that results from projections given by angles $\alpha$ and $\gamma$ that produce $i=65^{\circ}\pm12.5^{\circ}$. In the following we will focus on one of the two simulations of Section~\ref{sec:xovo} where $\thetaddo=30$ and $\thetaR=60^{\circ}$ and $\vyini=45\kms$ which, among the cases explored, reproduces most of the properties of DDO 68. The ICs (discs inclinations and sense of rotation) of the simulation are shown by red curves in Fig.~\ref{fig:ics} and summarized in Table~\ref{tab:ddo68sim}.

\subsection{The mock DDO 68}
\label{sec:mockddo}

The sets of panels of Fig.~\ref{fig:orbitmodel} show stellar surface brightness maps (left) and gas surface densities (right) for our fiducial simulation of DDO 68 ($\thetaddo=30^{\circ}$, $\thetaR=60^{\circ}$, $\vyini=45\kms$). The time of the simulation increases from top to bottom, the system is projected along the line-of-sight given by the triplet $(\alpha,\gamma,\beta)=(30^{\circ},225^{\circ},-20^{\circ})$, and Satellite S has been included in the simulation. We show the orbits of Satellite B and Satellite S constructed joining the centres of mass of the corresponding dark matter halos from subsequent snapshots using the shrinking sphere method \citep{Power2003}. The bottom panels show the maps corresponding to the time of the simulation that produces a configuration in agreement with observations. 
To guide the eye in the comparison with observations, the simulated stellar map in the bottom left panel of Fig.~\ref{fig:orbitmodel} is also shown in the right-hand panel of Fig.~\ref{fig:ddo68}, where we have indicated all the noteworthy structural features reproduced.

In each panel of Fig.~\ref{fig:orbitmodel}, the right is aligned with the north by the angle $\beta$, fixed requiring that the PA of DDO 68 falls within $19^{\circ}\pm10^{\circ}$, while the PA of Satellite S within $36^{\circ}{}^{+10^{\circ}}_{-18^{\circ}}$ \citepalias{Annibali2019b}. \citetalias{Cannon2014} do not provide errors on the PA of DDO 68; here we assume $10^{\circ}$ uncertainty. We have estimated the PA and the semi-minor ($c$) to semi-major ($a$) axis ratio of DDO 68 in the simulation computing the inertia tensor of the surface brightness map in Fig.~\ref{fig:orbitmodel} in an ellipse with a semi-major axis $4\kpc$ long, centred on the map brightest peak (for details see also \citealt{Lau2012}). The PA of DDO 68 is $\simeq26^{\circ}$, while Satellite S forms an angle of $\simeq45^{\circ}$ with the north. The ICs of Satellite S are listed in Table~\ref{tab:icsbestsim}.

Given the low resolution of the $\HI$ gas map provided by \citetalias{Cannon2014} ($\simeq 15\asec=0.9\kpc$ FWHM), we do not show gas maps downgraded to the resolution of observations but we keep a relatively high resolution for a more quantitative analysis of the simulation: we bin the gas particles into pixels $0.68\asec\times0.68\asec$ wide and smooth the resulting maps with a $1.9\asec$ gaussian FWHM. After $t=1\Gyr$ the projected distribution of the visible stellar component is very similar to the observed configuration of DDO 68 (see Fig.~\ref{fig:ddo68}). The satellite galaxies are disrupted by the gravitational force field of DDO 68 and, after less than $1\Gyr$, the stellar disc of Satellite B has almost lost any sign of coherent differential rotation, becoming, at $t\simeq1\Gyr$, the elongated stellar and gas stream that wraps the center of DDO 68. This stellar stream resembles DDO 68 B from the Tail to almost the Head, for approximately $10\kpc$, and it is very similar to configuration 1 in \citetalias{Sacchi2016}. As a slight mismatch, we note, however, that in DDO 68 the Cometary Tail is wider than in the outcome of the simulation, and that the Cometary Head is to the north in DDO 68, while in our simulation it forms to the north-east. Although it is reasonable to speculate that, for instance, a more extended stellar disc in Satellite B may produce configurations where the Cometeray Tail is spatially more similar to observations, for sake of simplicity we decided to not explore different galaxy models. The densest gas region in the simulation is offset with respect to the kinematic center of DDO 68. This is in agreement with observations, where the $\HI$ is denser on the eastern edge of DDO 68, where the Tail attaches to DDO 68, and it is filamentary and sparse on the western edge. The high density gas is spatially located where the remnants of the accreted Satellite B impact with DDO 68. In this very same portion, the galaxy hosts some of its brightest $\HII$ regions (\citealt{Pustilnik2005}; \citetalias{Cannon2014};\citetalias{Annibali2016};\citetalias{Annibali2019}). After $1\Gyr$ Satellite B retains a considerable fraction of gas, $\simeq2.5\times10^7\Msun$, about 40\% of its initial mass. The spacing between the star forming $\HII$ regions in DDO 68 ranges between $0.25\kpc$ and $0.75\kpc$. These distances are expected to be of the order of the Jeans length of the gas and thus contain information on its temperature and density. In the simulation, the gas forms dense arc-like agglomerates mostly distributed in the regions of the Head, the centre of DDO 68, and where Satellite B has recently crossed. The average separation between these agglomerates is $1-2\kpc$, comparable to the Jeans length computed in these very same regions in the simulation. The difference between the two values ($0.25\kpc-0.75\kpc$ in the observation, $1-2\kpc$ in the simulation) could be attributed to the fact that the simulations do not consider radiative cooling: the inability of the gas to cool makes the typical Jeans length larger. As a further check, we have also verified that the Jeans scale length computed where the dense agglomerates form in the simulations is larger than the characteristic size of the gas cells that populate that regions. This ensures that the gas resolution is enough to follow to process of gas fragmentation. 

We found that the presence of gas in the Tail constrains the orbit of its progenitor galaxy: if the orbital spin of Satellite B is highly misaligned with respect to that of the disc of DDO 68 (e.g. $\thetaddo=\thetaR=0^{\circ}$ and $\vy=-45\kms$ or, in general, most of the simulations with negative $\vy$) the average relative velocity between Satellite B and DDO 68 can be as high as $90\kms$. This high relative velocity makes gas ram-pressure particularly efficient in stripping the gas from Satellite B when it moves within or crosses the gas disc of DDO 68. Also, based on the simulations explored, we do not find any particular dependence of the outcome of the simulations on the disc spin of Satellite B: simulations with a clockwise or anticlockwise rotation of the disc of Satellite B produce very similar spatial distribution of the stellar and gaseous components.


The Tail is highly populated by young stars, in the age range $50-300\Myr$ (\citetalias{Annibali2016}; \citetalias{Sacchi2016}), compatible with the bulk of the star-formation activity observed in the whole system. The simulations neglect star formation and thus, by construction, stars in the Tail (Satellite B) trace the distribution of stars as old as $\simeq1\Gyr$. For the chosen mass-to-light ratios, the luminosity of Satellite B at $t\simeq1\Gyr$ is compatible with the luminosity of DDO 68 B. This implies a stellar mass of $\simeq1.1\times10^7\Msun$, meaning that only 10\% of the initial mass of Satellite B is stripped or, in the mock Fig.~\ref{fig:ddo68}, is fainter than the surface brightness limit of 29 mag$\asec^{-2}$. In the case of Satellite S, instead, its luminosity is compatible with that of stream S1 for the chosen mass-to-light ratio, implying a \lapex visible\rapex stellar mass of $1.56\times10^6\Msun$, about 40\% of the initial mass, in agreement with the estimate of \citetalias{Annibali2019b} of $(1.27\pm0.41)\times10^6\Msun$ and that thus motivates the use of a relatively larger initial stellar mass. 

A by-product of the interaction between Satellite B and DDO 68 is the formation of a low-luminosity overdensity of stars to the west (Fig.~\ref{fig:ddo68}) of DDO 68. The structure is similar in shape to the Arc, and it extends for a projected distance of $5\kpc$, as in the observations. The arc in the simulation is formed after Satellite B crosses the relevant region populated by the arc's stars, approximately $200-300\Myr$ before the system has reached a configuration similar to observations (middle and bottom left panels of Fig.~\ref{fig:orbitmodel}) and this very same interaction also induces the formation of a mild spiral pattern within the main galaxy. The time of this latest passage is perfectly comparable with the age of the youngest stars that populate the Arc ($200-300\Myr$ old), which strengthens the hypothesis that this episode of star formation was induced by the interaction between the Tail's progenitor and DDO 68. Also, comparing the same simulations differing only for the presence of  Satellite S, we can robustly conclude that Satellite S plays no role in the arc's formation, as observationally suggested by their very different stellar populations (\citetalias{Annibali2016}; \citealt{Annibali2019b}). 

\begin{figure*}
    \centering
    \includegraphics[width=1\hsize]{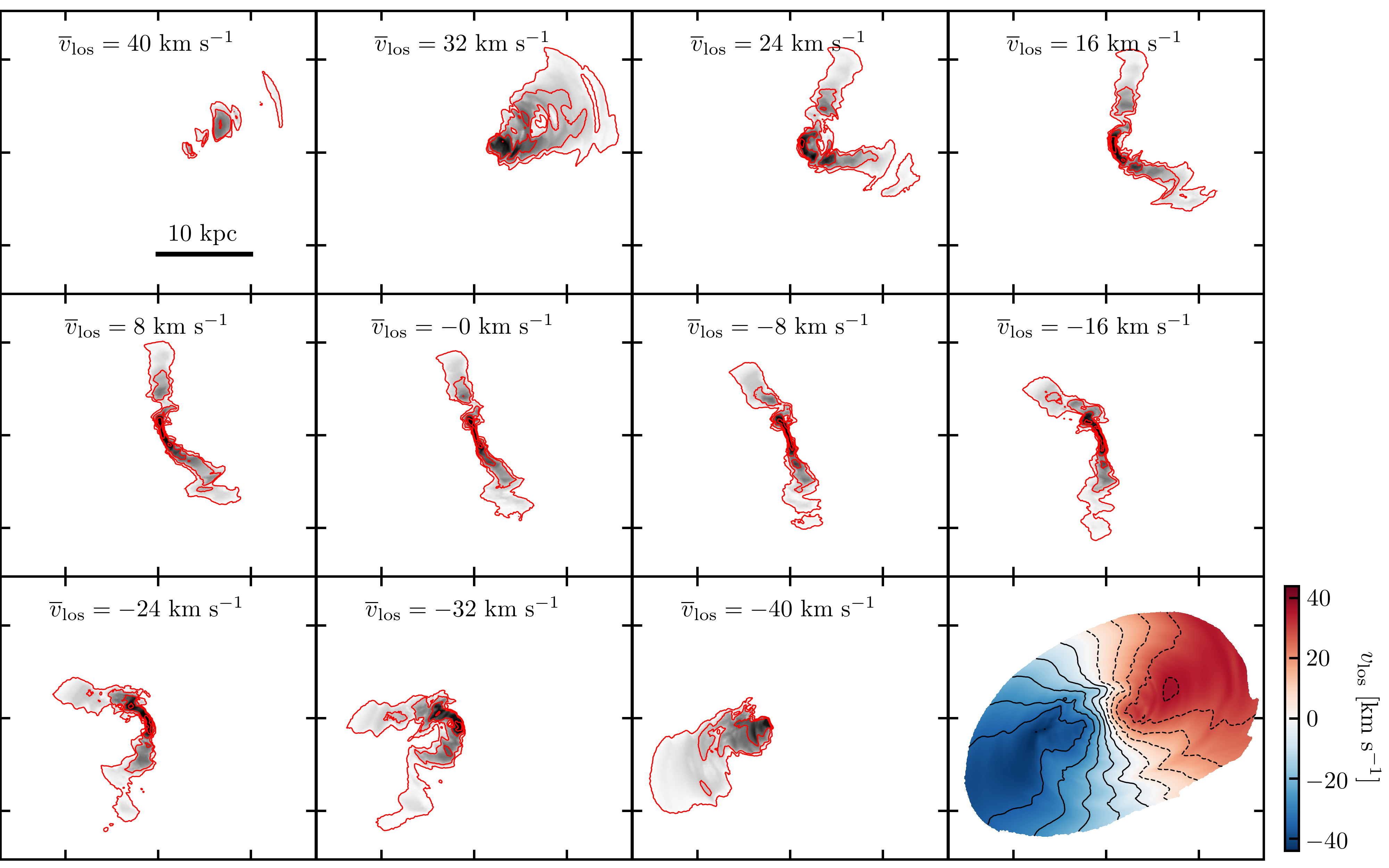}
    \caption{Maps of velocity channels showing the gas distribution of DDO 68 for the reference simulation of Section~\ref{sec:res}. Each channel is $8\kms$ wide, the average channel width is reported in each panel and the red contours mark column densities corresponding to $0.5\times10^{20}$, $1.26\times10^{20}$, $3.17\times10^{20}$, $8\times10^{20} \cm^{-2}$. The system is projected as in Fig.~\ref{fig:orbitmodel}, the pixel size $0.68\asec\times0.68\asec$ and the maps have been smoothed with a $1.9\asec$ FWHM beam, as in Fig.~\ref{fig:orbitmodel}. The bottom right panel shows the overall gas mass weighted line-of-sight velocity field: blue colors and negative velocities correspond to the disc approaching arm, while red colors and positive velocities to the disc receding arm. The velocity field is shown down to column densities of $0.5\times10^{20}\cm^{-2}$.}\label{fig:vfield}
\end{figure*}

In Fig.~\ref{fig:vfield} we show the gas distribution in channels of velocities for the snapshot corresponding to $t=1\Gyr$ (bottom-right panel of Fig.~\ref{fig:orbitmodel}). The maps resolution is the same as in the right-hand panels of Fig.~\ref{fig:orbitmodel}. Positive velocities show the disc receding arm and negative velocities the disc approaching arm. The overall observed velocity field is well reproduced by the simulation: the disc receding arm of DDO 68 is to the north and the disc has the same velocity gradient of $60-70\kms$ measured by \citetalias{Cannon2014} from the $\HI$ kinematics, and by \citetalias{Annibali2019} from the line-of-sight velocities of $\HII$ regions in DDO 68 B (for details see also Section~\ref{sec:met}). As a further comparison, in the bottom-right panel of Fig.~\ref{fig:vfield} we show the overall gas mass weighted line-of-sight velocity field from the simulation: the spatial resolution is the same as in the other panels, with red colors corresponding to the disc receding arm and blue colors to the approaching arm. The interaction with Satellite B does not provide any substantial perturbation to the gas kinematics of DDO 68, which keeps its coherent differential rotation about its symmetry axis. Although the velocity map of Fig.~\ref{fig:vfield} shows some irregularities caused by the interaction with Satellite B, when downgraded to the resolution of observations it is even more regular than that from \citetalias{Cannon2014} which shows, instead, typical kinks in the isovelocity contours associated to radial motions even at a resolution of $15\asec$. However, it is important to bear in mind that the simulations are dissipationless, so features in the gas distribution and velocity field possibly associated to star formation or SNe explosions in DDO 68 cannot be reproduced.

\section{The metallicity of DDO 68}
\label{sec:met}

\subsection{Observations}

\begin{figure*}
    \centering
    \includegraphics[width=1\hsize]{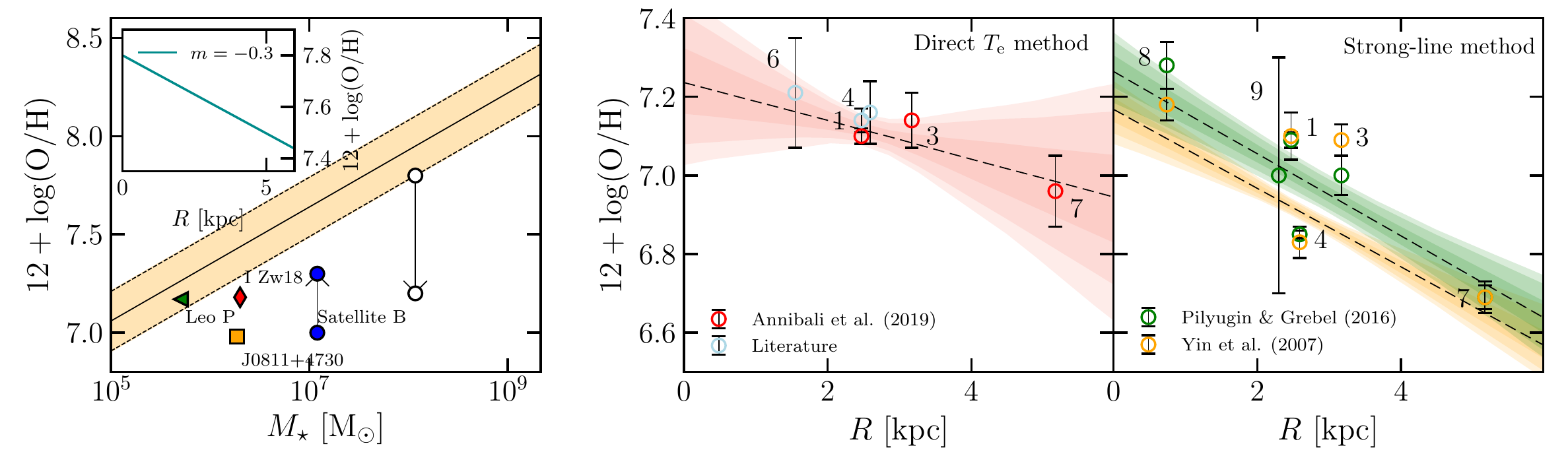}
    \caption{Left panel: MZR relation from \citet{Berg2012}. The light orange band shows the intrinsic scatter of the relation while the arrow between the white points starts from the metallicity of DDO 68 in the ICs of the simulation of Section~\ref{sec:met} and points at the average measured value. The black arrow between the blue circles starts from the initial oxygen abundance assumed for Satellite B in the simulation and points at the diluted metallicity. The green triangle, yellow square and the red diamond show the position in the mass-metallicity plane of Leo P, J0811+4730 and I Zw 18 respectively. The small inset shows the oxygen abundance as a function of the galactocentric distance (equation~\ref{for:ohvsr}) in the ICs of the simulation. Middle panel: $12+\logoh$ from six $\HII$ regions of DDO 68 measured with the direct-$\Te$ method as a function of the galactocentric distance. The red and cyan dots are measurements from \citetalias{Annibali2019}, and \citet{Izotov2007,Izotov2009}, respectively. We fit the measurements with relation \ref{for:ohvsr} and report the fiducial model (black dashed curve) and the $n\sigma$ uncertanties, with $n=1,2,3$ (light red bands). Right panel: same as the middle panel but showing measurements of the six $\HII$ regions from \citetalias{Annibali2019} calibrated using strong-line relations from \citet[green dots]{Pilyugin2016} and \citet[yellow dots]{Yin2007}, alongside the best fit models. The numbers in the middle- and right-hand panels refer to $\HII$ regions in DDO 68 according to the nomenclature of \citetalias{Annibali2019} (see Fig.~\ref{fig:ddo68}). Details on the fits are given in Table~\ref{tab:met}.}\label{fig:mzr}
\end{figure*}

\citetalias{Annibali2019} have recently provided new estimates of the metallicity of DDO 68 using deep spectra of six $\HII$ regions acquired with the Multi Object Double Spectrograph (MODS) at the LBT in the wavelength range $3500-10000\AA$. The authors show that the oxygen abundance can be as low as $12+\logoh=6.69\pm0.04$ in $\HII$ regions located in the Tail and up to $12+\logoh=7.28\pm0.06$ for $\HII$ regions in the Cometary Head. While these measures are based on strong-line method calibrations, \citetalias{Annibali2019} also provide independent estimates for three out of six $\HII$ regions based on the more robust direct-$\Te$ method which, combined with the ones from \cite{Izotov2007,Izotov2009}, give a slightly higher oxygen abundances of $12+\logoh=7.21\pm0.07$ for $\HII$ regions in the Head and $12+\logoh=6.96\pm0.09$ for $\HII$ regions in the Tail. In the middle- and right-hand panels of Fig.~\ref{fig:mzr} we show the direct-$\Te$ measures by \cite{Izotov2007,Izotov2009} and \citetalias{Annibali2019}, and the strong-line method measures by \citetalias{Annibali2019} as a function of the projected galactocentric distance (see also Fig. 11 of \citetalias{Annibali2016}). Although slightly different from each other, both methods confirm the metal deficiency of DDO 68 as well as the presence of a metallicity gradient.

\subsection{Dilution hypothesis}

Since the dilution of pristine gas recently accreted is one of the scenarios called upon to explain the existence of XMP-galaxies \citep{Spitoni2015}, motivated by the merger history of DDO 68 we explore to what extent it is realistic to speculate that a primeval gas distributed in Satellite B can affect the metal content of DDO 68 and vice versa. We note that in our hydrodynamical $N$-body models Satellite B accounts for only 6\% of the total gas mass budget, so a full dilution from $12+\logoh\simeq7.9$ (a value expected from the MZR as we will shortly discuss) to the average $12+\logoh\simeq7.2$ would be an unlikely expectation. For example, if we assume, for simplicity, a constant metallicity throughout DDO 68, zero metals in the accreted satellite, and efficient gas mixing over short time scales, the gas of Satellite B would lower the overall metallicity of DDO 68 by at most 0.05 dex, while a complete dilution to $12+\logoh\simeq7.2$ would only be achieved if Satellite B contributes with an unrealistic 75\% to the total gas mass budget. In DDO 68, however, most of the $\HII$ regions that provide measures of oxygen abundances are located along the Cometary Tail. According to our hydrodynamical $N$-body models, after $1\Gyr$ Satellite B retains a considerable fraction of its initial gas (see Fig.~\ref{fig:orbitmodel}), and this gas, similarly to what is observed, is displaced mostly along the structure that forms the Tail and can thus fuel the formation of new stars there. If coupled with an inefficient gas mixing, such feature makes us wonder if the $\HII$ regions of DDO 68 may be made up by the low-metallicity gas of the satellite galaxy and, then, if the extreme metal deficiency of DDO 68 is, instead, the result of an incorrect classification that attributes to a galaxy with stellar mass $\simeq10^8\Msun$ (DDO 68), the oxygen abundance of a galaxy with a lower stellar mass $\simeq10^7\Msun$ (the Tail).

\subsubsection{Simulation set-up}

To test this hypothesis, we attribute to DDO 68 an initial average oxygen abundance consistent with that expected from the MZR given the galaxy current stellar mass, while Satellite B is given an initial oxygen abundance typical of XMP galaxies of similar stellar mass ($10^7\Msun$). We let the metallicity field passively evolve in the simulations and study whether gas mixing can, or cannot, efficiently pollute the metallicity of Satellite B.

We rely on the MZR from \cite{Berg2012} which, given $\Mstar=1.17\times10^8\Msun$, implies $12+\logoh=7.95\pm0.15$ and  we take its lower limit $12+\logoh=7.8$ as initial average metallicity of DDO 68 in the simulations. We allow the initial metallicity of DDO 68 to decrease inside-out according to the relation (\citealt{Sanchez2014,Magrini2016}; \citetalias{Annibali2019})
\begin{equation}\label{for:ohvsr}
    [12+\logoh](R) = \Ao + m\biggl(\frac{R}{\Riiv}\biggr)
\end{equation}
where $\Riiv=4.95\kpc$ is the isophotal radius \citepalias{Annibali2019}, $m$ is the slope of the relation and $\Ao\equiv[12+\logoh](0)$ its central metallicity. We set $\Ao =7.8$ and consider the case of a steep gradient $m=-0.3$. We set the oxygen abundance of Satellite B to $12+\logoh=7$ (constant throughout the galaxy), a value consistent with that of XMP galaxies with similar stellar mass. We note that the oxygen abundance extrapolated from the MZR for a galaxy with stellar mass $\simeq10^7\Msun$ would be $7.64\pm0.15$, higher than the value measured in DDO 68. In the left-hand panel of Fig.~\ref{fig:mzr} we show the actual and expected positions of DDO 68 in the MZR, the oxygen abundance of Satellite B, together with the metallicity of some XMP galaxies with stellar mass similar to that of Satellite B, while in the small inset we show the initial metallicity distribution of DDO 68 (\ref{for:ohvsr}) in the simulation.


The simulations do not model cooling nor star formation, so the metallicity field passively evolves without affecting any of the system's structural and kinematic properties according to the continuity equation
\begin{equation}\label{for:cont}
    \frac{\DD(Z\rhogas)}{\DD t} + \nabla\cdot(Z\rhogas \vecv) = 0,
\end{equation}
which is related to the log-scale oxygen abundance by
\begin{equation}\label{for:ox}
Z = 10^{[12 +\log(\OH)] - A_{\odot}}\Zsun,
\end{equation}
with $A_{\odot}\equiv12 +\log([\OH]_{\odot})= 8.76$ the solar oxygen abundance \citep{Caffau2008}. Note that with equation~(\ref{for:ox}) we implicitly assume, for simplicity, that only oxygen contributes to the gas metal composition. However, this assumption does not represent a limitation of the model, given the linearity of equation (\ref{for:cont}). Actual chemo-dynamical evolution models should be computed to follow other chemical species; this is, however, beyond the scope of this paper.


\subsubsection{Results}

\begin{figure}
    \centering
    \includegraphics[width=.8\hsize]{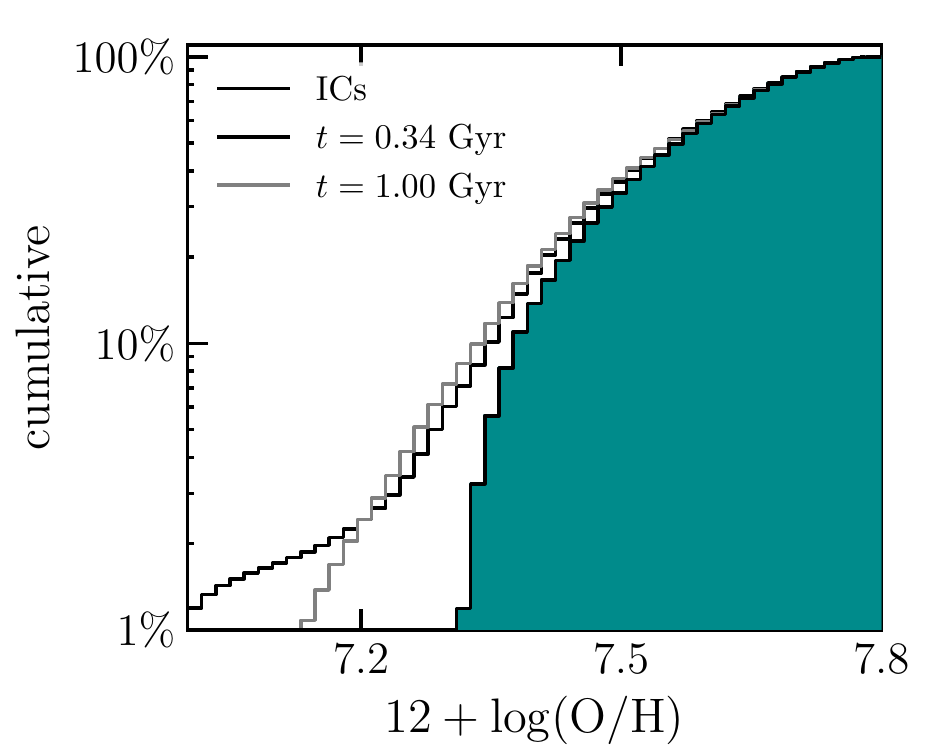}
    \caption{Cumulative, gas-mass weighted metallicity distribution as a function of time for the simulation of Section~\ref{sec:met}. The curves correspond to distributions computed from the ICs (dark cyan), after $t=340\Myr$ (black curve) and after $t=1\Gyr$ (grey curve).}\label{fig:metcum}
\end{figure}

In Fig.~\ref{fig:metcum} we show a sequence of histograms representing the time evolution of the cumulative, gas-mass weighted, metallicity distribution in the simulation of Section~\ref{sec:met}. Different curves refer to different snapshots: the ICs are dark cyan, the black line corresponds to $t=340\Myr$, and grey is for $t=1\Gyr$, the time after which the system produces a configuration similar to the observed one. To highlight the effects of the merger between Satellite B and DDO 68, we have considered only gas particles $8\kpc$ away from the centre of DDO 68 to build the cumulative metallicity distributions. The impact enhances the total amount of gas within $8\kpc$, fueling the low metallicity tail of the cumulative distributions. After $t=340\Myr$, more than 1\% of the gas within $8\kpc$ is below $12+\logoh=7$, while after $1\Gyr$ more than 2\% still has metallicity as low as 7.2, close to the average oxygen abundance measured in the $\HII$ regions of DDO 68.

\begin{figure*}
    \centering
    \includegraphics[width=1\hsize]{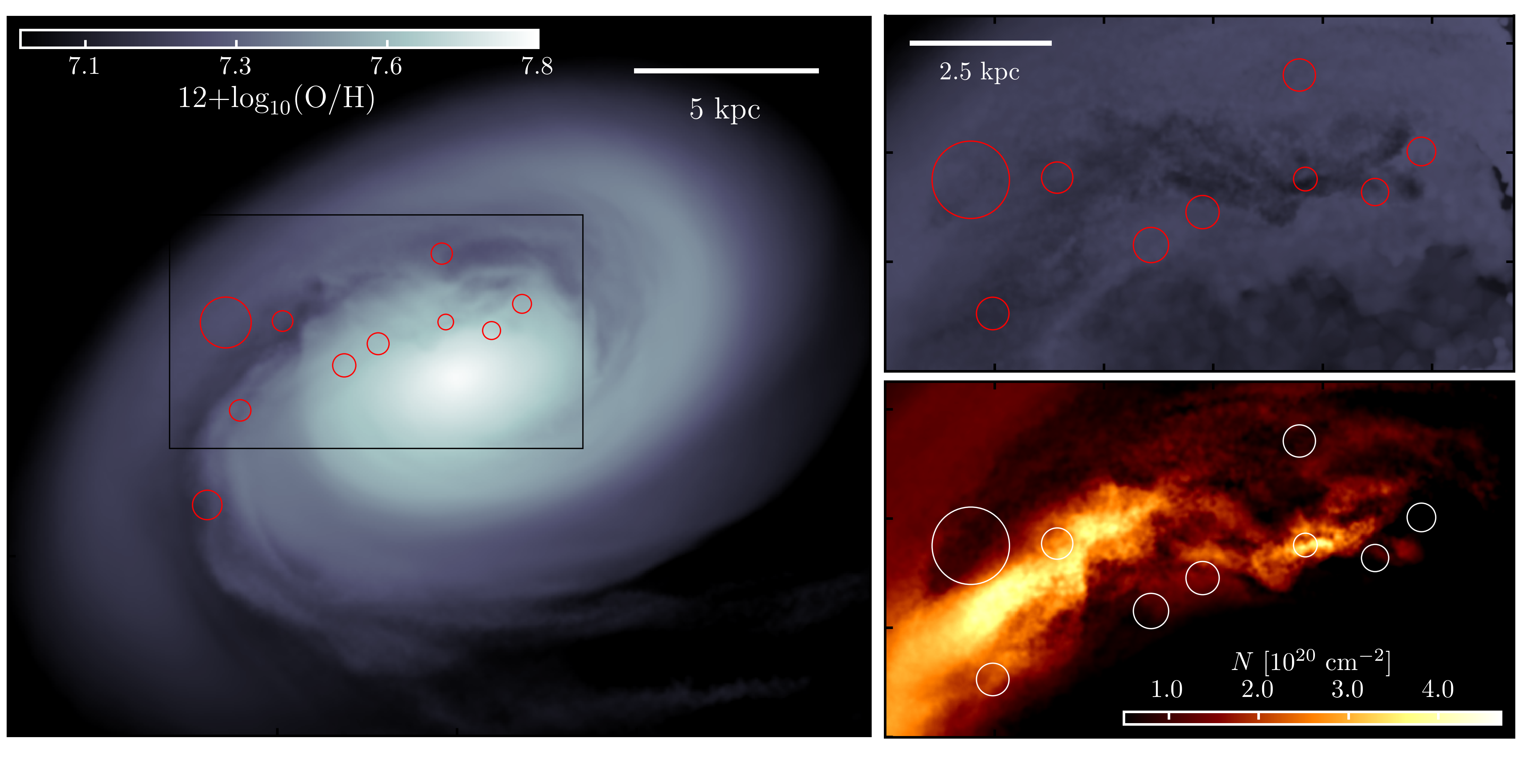}
    \caption{Left panel: gas mass-weighted metallicity map from the simulation of Section~\ref{sec:met} with $m=-0.3$. The line-of-sight is the same as in Fig.s~\ref{fig:orbitmodel} and ~\ref{fig:vfield} and the simulation has evolved for $t=1\Gyr$ (bottom panels of Fig.~\ref{fig:orbitmodel}). Top-right panel: zoom-in view of the region marked with a black box in the left panel. To highlight substructures in the gas of Satellite B, the metallicity map has been obtained considering only gas particles with oxygen abundance $12+\logoh<7.3$ (see right panel of Fig.~\ref{fig:metcum}). Bottom-right panel: gas surface density of the weighted metallicity map shown in the top-right panel. The red and white circles in all panels show the regions of extractions used to compute the local oxygen abundances shown in Fig.\ref{fig:simdist}.}\label{fig:simmet}
\end{figure*}

In the left panel of Fig.~\ref{fig:simmet}, we show, instead, the gas-mass weighted metallicity map after $t=1\Gyr$, assuming the same line-of-sight and resolution of the right panels of Fig.~\ref{fig:orbitmodel} or Fig.~\ref{fig:vfield}. The crossing and the effect of Satellite B is evident looking at the darker region close to the galaxy centre: Satellite B produces a low metallicity component within the high metallicity gas of DDO 68 that wraps the galaxy centre along a path more or less consistent with its orbit. To highlight the contribution of Satellite B, in the top right panel we show a zoom-in view of the region marked by a black rectangle in the left panel, obtained considering only gas particles with oxygen abundance $12+\logoh<7.3$, while the bottom-right panel of Fig.~\ref{fig:simmet} shows the corresponding projected particle number density map: note how a considerable fraction of the low-metallicity gas follows the optical counterpart, fills the Tail and the eastern edge of DDO 68 in the simulation. 

\begin{figure}
    \centering
    \includegraphics[width=.9\hsize]{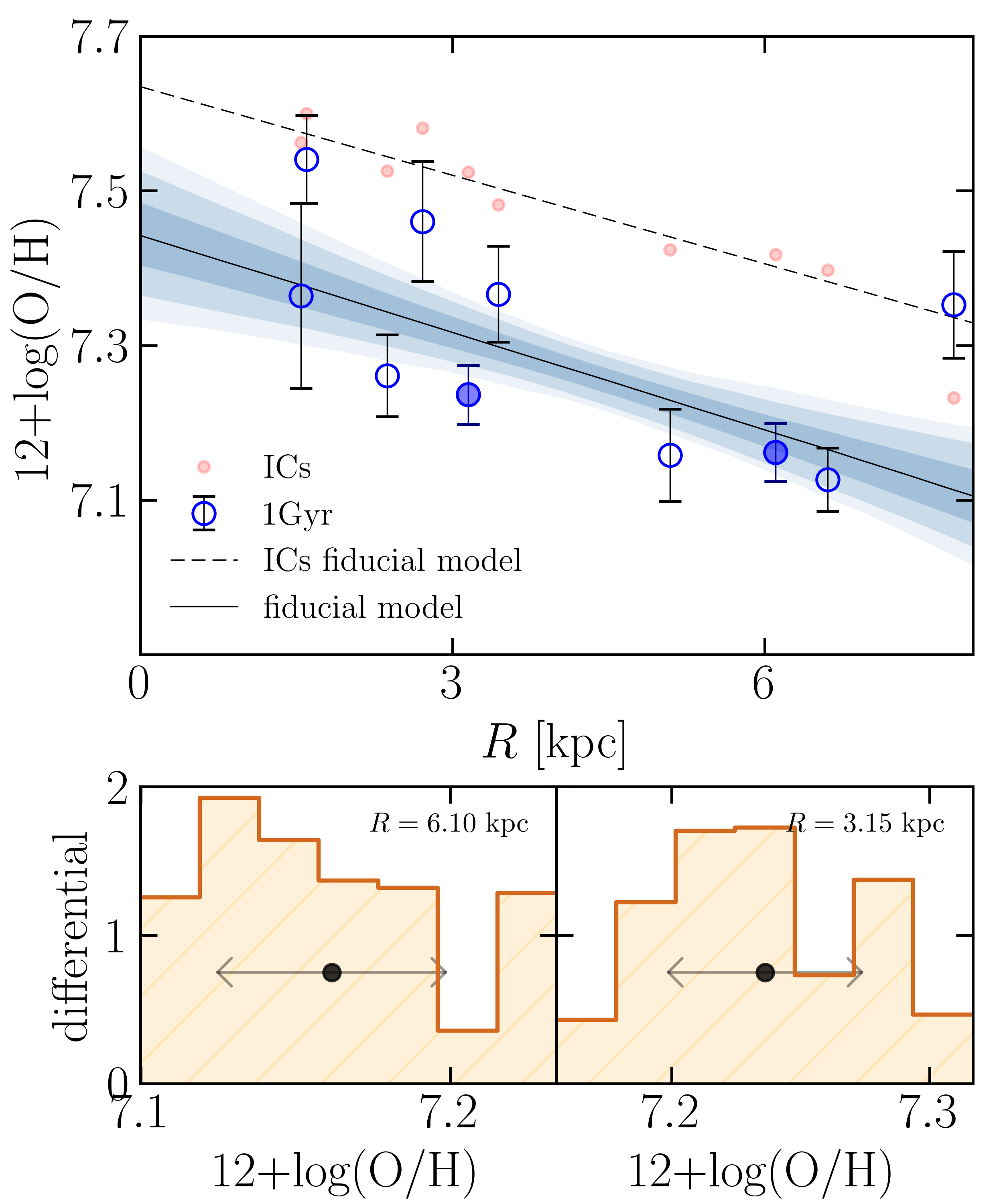}
    \caption{Top panel: oxygen abundance (dots with errorbars) as a function of the projected distance from the centre of the galaxy for the simulation with initial intrinsic metallicity gradient $m=-0.3$, after the simulation has evolved for $1\Gyr$. The abundances are computed in the spherical regions along Satellite B, as shown in Fig.~\ref{fig:simmet} (see text for details). The light red dots show the oxygen abundances computed in the same extraction regions but in the ICs. Each measure is plotted as $\mu\pm\sigma$, with $\mu$ and $\sigma$ the mean and dispersion of the gas-mass weighted metallicity distribution of that extraction region, respectively, converted into oxygen abundances. The dashed and solid black lines show the best fit relations \ref{for:ohvsr} derived from the evolved simulation and the ICs (best fit parameters are given in Table~\ref{tab:met}), respectively. Bottom panels: gas-mass weighted metallicity distribution for two extraction regions marked by full dots in the top panels. The black dots show the mean metallicity while the arrows a region $2\sigma$ wide.}
    \label{fig:simdist}
\end{figure}

Observationally, five $\HII$ regions are in the Head, while the remaining four are more or less distributed along the Cometary Tail. To make a quantitative comparison with \citetalias{Annibali2019}, we have derived the oxygen abundance of the galaxy in a selection of ten regions more or less located along the accreted satellite, where we reach the lowest metallicities after $t=1\Gyr$, and we plot them against the projected distance from the galaxy centre in the top panel of Fig.~\ref{fig:simdist}. All the regions are shown in Fig.~\ref{fig:simmet}: four of them are located along the trajectory of Satellite B (the rightmost circle in Fig.~\ref{fig:simmet} is the current center of mass of Satellite B), the remaining six have been selected within the disc of DDO 68 or where the gas of Satellite B has been stripped, such as to mimic approximately the spatial distribution of the $\HII$ regions observed in DDO 68 and to trace the metallicity field over regions affected the most by the crossing of Satellite B. Once the centres of the regions have been determined, we selected all gas particles within a sphere with a radius of $0.2\kpc$, if the number of gas particles is less than 50 we iteratively widen the radius of the extraction regions to contain at least 50 gas particles. We then compute the gas mass-weighted metallicity distribution in each region and take as a measure of metallicity and error the average and the dispersion of the distribution, respectively. Note that with the criterion adopted for mesh refinement (see Section \ref{sec:eq}), fixing the minimum number of particles approximately fixes the minimum gas mass sampled by each extraction region and that this requirement allows us to have a good sampling of the local gas-mass weighted metallicity distribution. Finally, $Z$ is converted in oxygen abundance using equation (\ref{for:ox}).

The mock observations of the oxygen abundance are shown in the top panel of Fig.~\ref{fig:simdist} together with the oxygen abundances computed in the same regions of the ICs to highlight the effect of dilution. In the bottom panels of Fig.~\ref{fig:simdist} we show, instead, the local gas-mass weighted metallicity distributions for two extraction regions used to compute the relations in the top panel. The dispersion of the gas-mass weighted metallicity distributions provides a quantitative indication of local gas-mixing efficiency: sharp-peaked distributions (thus with small dispersion) indicate well-determined oxygen abundances, broad distributions are associated to a mixture of relatively low and high gas metallicity coexisting in the same small spatial region. We fitted both sets of oxygen abundances (ICs and evolved simulation) with the relation (\ref{for:ohvsr}), as done with the observations shown in Fig.~\ref{fig:mzr}, and list the best fit results in Table~\ref{tab:met}. 

On average, the overall projected radial distribution of oxygen abundance decreases by $\simeq0.2$ dex, while we do not find any substantial change in its slope (see also Table~\ref{tab:met}), which is very similar to the slope inferred from \cite{Annibali2019} and \cite{Izotov2007,Izotov1999} data. Even if the contribution of Satellite B to the total gas mass is modest we find that gas mixing is not efficient enough to increase the low-metallicity of the satellite gas to values similar to those of the main galaxy. A considerable fraction of this gas keeps a relatively low metallicity. For instance, at $6\kpc$ from the galaxy centre we infer an oxygen abundance 7.1, only 0.1 dex higher than what is found from the direct-$\Te$ measures of \citetalias{Annibali2019} (see middle panel of Fig.~\ref{fig:mzr}). At this point, it must be emphasised that our experiment is intended to explore a favorable, though realistic, scenario that maximises, in principle, the chance to keep a relatively low metallicity in the regions populated by the gas of Satellite B. More complicated simulations including a detailed treatment of chemical evolution must be used to follow also other chemical species. We have indeed assumed that DDO 68 has the lowest initial metallicity possible from the MZR, that this value is measured in the galaxy centre and that the initial oxygen radial distribution has a relatively steep gradient to favor low oxygen abundances in the galaxy outer parts. Also, we have neglected star formation with its subsequent metal enrichment which, over $1\Gyr$ would give a non-negligible contribution given the measured star-formation activity. Nevertheless, we did not explore in our simulations scenarios with different relative gas contents in DDO 68 and Satellite B which could easily favor a larger gas mass for Satellite B (within a factor of two at least) making gas mixing even less efficient. Putting all these considerations together, and considering the current model limitations, it does seem plausible that the $\HII$ regions from which we infer the metal content attributed to DDO 68 are, instead, tracing the low metallicity of the accreted satellite.

\begin{table}
    \centering
    \caption{Parameters resulting from the fit of the oxygen abundance profiles of Fig.s~\ref{fig:mzr} and~\ref{fig:simdist} with model~(\ref{for:ohvsr}). We perform a chi-square fit and we estimate the best fitting parameters as the 50-th percentile of the one-dimensional posterior distributions, while the corresponding errors as the 16-th and 84-th percentiles.}
    \begin{tabular}{ccc}
    \hline\hline
      &  $m$ &  $\Ao$ \\
    \hline\hline
    ICs         &   $-0.190\pm0.001$            &   $7.635\pm0.007$ \\
    $1\Gyr$     &   $-0.208^{+0.044}_{-0.042}$  &   $7.443^{+0.041}_{-0.043}$ \\
    \hline\hline
    \citetalias{Annibali2019}   & $-0.251^{+0.158}_{-0.156}$ & $7.242^{+0.082}_{-0.085}$ \\
    \citet{Yin2007}             & $-0.495^{+0.051}_{-0.055}$ & $7.168^{+0.031}_{-0.030}$ \\
    \citet{Pilyugin2016}        & $-0.516^{+0.058}_{-0.062}$ & $7.264^{+0.041}_{-0.040}$ \\
    \hline\hline
    \end{tabular}
    \label{tab:met}
\end{table}

\section{Conclusions}
\label{sec:conc}

We have presented state-of-the-art hydrodynamical $N$-body simulations able to explain most of the peculiarities observed in DDO 68. In agreement with the previous findings of \citetalias{Annibali2016}, the optical shape of DDO 68 can be easily explained as the result of the interaction between three systems: a dominant galaxy, DDO 68, with a dynamical mass of the order of $10^{10}\Msun$ and two smaller satellite galaxies with masses 1/20 and 1/150 times the mass of DDO 68. The most massive satellite galaxy is the progenitor of the elongated stellar stream to the south of DDO 68, the Cometary Tail, while the second stellar stream, DDO 68-S1, is well explained as the remnant of the third galaxy. The arc-like structure to the west, the Arc, is also produced by the simulation as the result of the interaction between DDO 68 and the larger satellite, confirming their association as suggested by their stellar population. 

We produced mock observations to be compared with real observations and, differently from previous simulations of DDO 68, all the systems are represented with realistic galaxy models. Simulations include gas hydrodynamics, complemented by a passively evolving scalar field that traces the overall behavior of the metallicity field of DDO 68. The asymmetric gas distribution is qualitatively well reproduced: after the interaction the gas is more concentrated and dense towards the Cometary Tail rather than the galaxy centre. The interaction does not provide a disruptive perturbation to the velocity pattern of DDO 68 gas disc, and, after the interaction, the velocity gradient of $\simeq70\kms$ seen in $\HI$ \citepalias{Cannon2014} and $\HII$ regions data, and the position of the disc approaching and receding arms are reproduced.


Our study confirms that, in DDO 68, the irregularities observed throughout the galaxy are not the result of an interaction with a smaller companion as DDO 68 C, but we are more probably looking at a multiple accretion of smaller systems. According to cosmological simulations it is not surprising that galaxies as DDO 68 can host satellite galaxies of smaller mass \citep{Dooley2017,Patel2018,Bose2018}. Indeed, dwarf galaxies in isolation or in low-dense environments would be less susceptible to potential disruptive interactions such as those that dwarf galaxies of the same mass but close to more massive hosts would experience. In terms of mass and expected number of satellites, the small companions of DDO 68 are in perfect agreement with predictions from simulations \cite[e.g.][]{Dooley2017} and, more generally, satellites of satellites or associations of dwarf galaxies have been discovered also in several other systems \citep[see ][]{Bellazzini2013,Erkal2020,Carlin2021,MartinezDelgado2021}, 

We have investigated the possibility that the $\HII$ regions of DDO 68, mostly located along the Cometary Tail and the only viable tool to measure the galaxy metallicity, actually trace the oxygen abundance of the accreted satellite whose lower metallicity gas has been inefficiently mixed, rather than that of DDO 68. We attributed to DDO 68 an initial metallicity as expected from the MZR and to the satellite galaxy an average metallicity similar to that of dwarf galaxies with similar mass in the mass-metallicity plane. We followed the time evolution of the overall metallicity field and found that during the interaction between the two galaxies, gas mixing is inefficient and that, along the Cometary Tail, the gas metallicity keeps relatively low and very close to the values measured in DDO 68. According to our model, it is likely that the large offset in the MZR can be explained as the result of a misclassification, since the $\HII$ regions in DDO 68 may be tracing the metallicity of its $10^7\Msun$ accreated galaxy and not of the main body at $\simeq10^8\Msun$. This would bring DDO 68 position in a region of the mass-metallicity plane populated by other XMP galaxies. The only $\HII$ region that in DDO 68 may be tracing the actual galaxy metallicity is region 8 (see Fig.~\ref{fig:ddo68}) which has, indeed, the largest metallicity ($12+\logoh\simeq7.3$, see right-hand panel Fig.~\ref{fig:mzr}). Also, as pointed out by \citetalias{Annibali2019}, region 8 has the most uncertain measures of oxygen abundances that only rely on strong-line methods, so estimates higher of few dex, that would be more compatible with the value expected on the basis of the MZR, cannot be excluded.


\section*{ACKNOWLEDGMENTS}
We acknowledge the use of computational resources from the parallel computing cluster of the Open Physics Hub (\href{https://site.unibo.it/openphysicshub/en}{https://site.unibo.it/openphysicshub/en}) at the Physics and Astronomy Department of the University of Bologna. We acknowledge funding from the INAF Main Stream program SSH 1.05.01.86.28. FM is supported by the Program ‘Rita Levi Montalcini’ of the Italian MIUR. MC acknowledges the support of INFN \lapex Iniziativa specifica TAsP\rapex. RP warmly thanks G. Sabatini for useful suggestions and comments. 

\section*{Data availability}

The version of the Arepo code used to carry out the simulations analysed in this work is publicly available at \href{https://arepo-code.org/}{https://arepo-code.org/}. All the simulations used in this article will be shared on request to the corresponding author.

\bibliography{paper}
\bibliographystyle{mnras}

\appendix

\section{Comparison with the simulation of A16}
\label{app:B}

\begin{figure*}
    \centering
    \includegraphics[width=1\hsize]{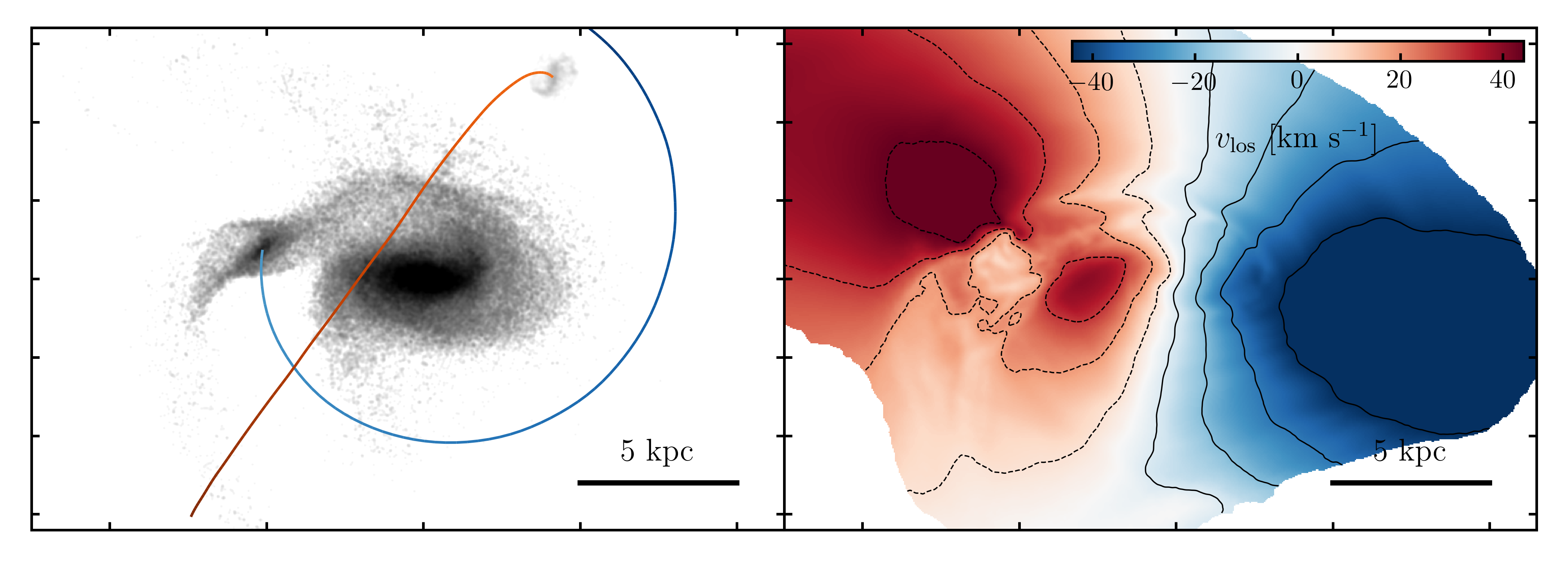}
    \caption{Left panel: stellar surface brightness map from the simulation of \ref{app:B}. The line-of-sight is given by $(\alpha,\gamma,\beta)= (-75^{\circ},180^{\circ},-20^{\circ})$, while the map resolution is the same as in Fig.~\ref{fig:orbitmodel}. The mass-to-light ratios attributed to the particles of DDO 68, Satellite B and Satellite S are $\MLgA=0.51$, $\MLgB=0.51$ and $\MLgS=4$, respectively. The blue and orange lines show the orbits of Satellite Band Satellite S. Right panel: gas-mass weighted line-of-sight velocity field. Red and blue correspond to the disc approaching and receding arm, respectively. The map resolution is the same as in the left panels of Fig.~\ref{fig:orbitmodel} or Fig.~\ref{fig:vfield}.}\label{fig:nipoti}
\end{figure*}

For a quantitative comparison and to test the need of new simulations, we produced a new version of the simulation of \citetalias{Annibali2016} that includes a gaseous component in DDO 68 and DDO 68 B, and accounts for discs thickness and self-gravity of stars and gas. The galaxy models employed for DDO 68 and satellite S are the same as the ones of Section~\ref{sec:freeparam}, while Satellite B\footnote{Note that in the nomenclature of \citetalias{Annibali2016} our Satellite B is their Satellite T.} in \citetalias{Annibali2016} has twice as dark-matter mass than ours. We have then re-sampled the ICs of Satellite B, doubling its dark-matter mass, and we have re-run the simulation in isolation to let the system shift towards equilibrium, as done in Section (\ref{sec:eq}). The initial phase-space position of Satellite B is the same as ours, while the initial position of Satellite S is at $\xyzi=(-10\kpc, 4.37\kpc,0)$, with initial velocity $\vxyzi=(0, 0, 45\kms)$. The disc of DDO 68 is inclined by $\thetaddo=45^{\circ}$ (using the same formalism of Section~\ref{sec:xovo}), while the discs of Satellite B and S in the ICs lie on the $xy$ plane. The orbital spin of Satellite B, and the discs spin of Satellite B and Satellite S are aligned and are all parallel to the negative $z$-axis (clockwise rotation). The disc of DDO 68 rotate such that its spin forms a $45^{\circ}$ with respect to the negative $z$-axis. The mass resolution and time duration of the simulation are the same as the simulations of Section~\ref{sec:eq}.

In Fig.~\ref{fig:nipoti} we show the stellar surface density map (left) and gas mass weighted velocity field (right), obtained after the simulation has evolved for $t\simeq0.54\Gyr$, projected along a line-of-sight given by the angles $(\alpha,\gamma,\beta)= (-75^{\circ},180^{\circ},-20^{\circ})$, as in \citetalias{Annibali2016}. For stars, we adopted the same resolution and surface brightness limit of the left-hand panel of Fig.~\ref{fig:orbitmodel}, while for gas the same as in Fig.~\ref{fig:vfield}. Differently from the simulation shown by \citetalias{Annibali2016}, since stars are not mass-less particles, we converted mass into luminosities adopting $\MLgA=\MLgB=0.51$ as mass-to-light ratios for DDO 68 and Satellite B, and $\MLgS=4$ for Satellite S. The addition of gas hydrodynamics and more complicated galaxy models do not sensibly change the resulting surface brightness map, which shows approximately the same structure as in the old simulation: induction of mild spiral pattern, formation of a low-surface brightness structure to the west (Arc) and, most importantly, the formation of the Cometary Tail from Satellite B as an elongated structure to the south of DDO 68, similar to configuration 2 of \citetalias{Sacchi2016}. We are not able, however, to reproduce gas kinematics: as shown in the right-hand panel, the disc receding arm is to the south rather than the north as it should be. We also run a simulation where the disc of DDO 68 rotates in the opposite direction, but, in this case, the optical structure is very different from observation for the selected (and for any) line-of-sight. This experiment highlights the complexity of reproducing all the kinematic and structural constraints at once and motivates why we preferred to systematically explore a large portion of the parameter space, focusing on those free parameters that regulate the relative direction of the orbital spin of Satellite B and discs spins of DDO 68 and Satellite B.

\section{Computation of mass-to-light ratios}
\label{app:A}

DDO 68-S1 is dominated by very old stars ($>2\Gyr$; \citetalias{Annibali2016}, \citealt{Annibali2019b}). Stars as young as $20-300\Myr$ have been detected overall DDO 68 A and DDO 68 B, while only the former has also a significant fraction of stars with ages as old as $2\Gyr$ \citepalias{Sacchi2016}, so we expect their mass-to-light ratios to be very different from one another given their very heterogeneous stellar populations. We compute the mass-to-light ratios of DDO 68 A and DDO 68 B relyin on the LBT $g$ band (see Fig.s~\ref{fig:ddo68}) observations of \citetalias{Annibali2016} and assuming both configurations from \citetalias{Sacchi2016}: a scenario where DDO 68 B extends from the Cometary Tail to the Cometary Head (configuration 1), and a scenario where DDO 68 B is only made by  the Tail (configuration 2).

In the $g$ band, and in configuration 1, we measure an integrated magnitude $\magg=14.83\pm0.02$ for DDO 68 A and $\magg=17.33\pm0.2$ for DDO 68 B. 
The magnitudes have been corrected for the foreground reddening using E(B-V)=0.018. Assuming the solar absolute magnitude in the $g$ band $\Maggsun=5.23$ \citep{Willmer2018}, a distance $d=12.65\Mpc$ \citepalias{Sacchi2016}, and a total mass in stars of $\Mstar\simeq1.17\times10^8\Msun$ for DDO 68 A and $\Mstar\simeq1.2\times10^7\Msun$ for DDO 68 B (see also Table~\ref{tab:ddo68ics}), we get
\begin{equation}
    \MLgA\simeq0.51\pm0.01,\quad\MLgB\simeq0.52\pm0.10,
\end{equation}
i.e. the mass-to-light ratios in the $g$ band of DDO 68 A and DDO 68 B, respectively. 
A similar calculation in configuration 2 gives approximately the same mass-to-light ratios.


The integrated g-band magnitude of the stellar stream DDO 68-S1 is $\magg=22.08\pm0.4$. At the same distance of DDO 68 A and DDO 68 B, and given the very uncertain stellar mass $\Mstar=(1.27\pm0.41)\times10^6$ (\citealt{Annibali2019b}), we get a g-band mass-to-light ratio
\begin{equation}
    \MLgS\simeq4.5\pm2.1.
\end{equation}


\label{lastpage}

\end{document}